\documentclass[a4paper,11pt]{article}
\usepackage[utf8]{inputenc}
\usepackage[T1]{fontenc}
\usepackage[english]{babel}
\usepackage[hmargin={32mm,32mm},vmargin={32mm,35mm}]{geometry}
\usepackage{amsmath}
\usepackage{amsfonts}
\usepackage{amssymb}
\usepackage[mathscr]{euscript}
\usepackage{enumerate}
\usepackage{graphicx}
\usepackage{bbm}
\usepackage{color}
\usepackage{xspace}
\usepackage{url}
\usepackage[hidelinks]{hyperref}
\usepackage{caption}
\usepackage{enumitem}
\usepackage{bookmark}
\usepackage{csquotes}
\usepackage[
	bibstyle=phys,
	biblabel=brackets,
	citestyle=numeric-comp,
	sorting=none,
	doi=false,
	eprint=true,
	pageranges=false,
	maxbibnames=10,
	backend=biber
]{biblatex}

\DeclareFieldFormat[article,inproceedings,inbook]{title}{\mkbibitalic{#1\isdot}}
\AtEveryBibitem{
	\ifentrytype{book}{
		\clearfield{series}
		\clearfield{volume}
	}{}
	\ifentrytype{inbook}{
		\clearfield{pages}
	}{}
}

\newcommand{\bra}[1]{\langle #1 |}
\newcommand{\ket}[1]{|#1\rangle}

\newcommand{\eg}{e.g.\@\xspace}
\newcommand{\ie}{i.e.\@\xspace}
\newcommand{\Eq}{Eq.\@\xspace}
\newcommand{\Eqs}{Eqs.\@\xspace}
\newcommand{\Fig}{Fig.\@\xspace}

\newcommand{\updown}[2]{^{#1}_{\phantom{#1}#2}}
\newcommand{\downup}[2]{_{#1}^{\phantom{#1}#2}}

\newcommand{\Tr}{\operatorname{Tr}}
\newcommand{\D}{{\mathscr{D}}}

\newcommand{\Rt}{\,{}^{(3)}\!R}
\newcommand{\threedots}{{\lower.25em\hbox{\vdots}}}

\numberwithin{equation}{section}

\begin{document}

\begin{center}

\Large
\textbf{On the dynamics of single-vertex states in\\quantum-reduced loop gravity}

\vspace{16pt}

\large
Ilkka Mäkinen

\normalsize

\vspace{12pt}

National Centre for Nuclear Research \\
Pasteura 7, 02-093 Warsaw, Poland

\vspace{8pt}
 
ilkka.makinen@ncbj.gov.pl

\end{center}

\renewcommand{\abstractname}{\vspace{-\baselineskip}}

\begin{abstract}
	\noindent In this article we examine a Hamiltonian constraint operator governing the dynamics of simple quantum states, whose graph consists of a single six-valent vertex, in quantum-reduced loop gravity. To this end, we first derive the action of the Hamiltonian constraint on generic basis states in the Hilbert space of quantum-reduced loop gravity. Specializing to the example of the single-vertex states, we find that the Euclidean part of the Hamiltonian bears a close formal similarity to the Hamiltonian constraint of Bianchi I models in loop quantum cosmology. Extending the formal analogy to the Lorentzian part of the Hamiltonian suggests a possible modified definition of the Hamiltonian constraint for loop quantum cosmology, in which the Lorentzian part, corresponding to the scalar curvature of the spatial surfaces, is not assumed to be identically vanishing, and is represented by a non-trivial operator in the quantum theory.
\end{abstract}

\section{Introduction}

Quantum-reduced loop gravity is a simplified model of loop quantum gravity, which was introduced by Alesci and Cianfrani in \cite{Alesci:2012md, Alesci:2013xd} and was initially motivated as an attempt to study the cosmological sector of loop quantum gravity and to clarify the relation between loop quantum cosmology and full loop quantum gravity \cite{Alesci:2014rra, Alesci:2015nja, Alesci:2016gub}. The Hilbert space of quantum-reduced loop gravity is a specific subspace of the kinematical Hilbert space of loop quantum gravity. This space, which is spanned by the so-called reduced spin network states, can be obtained as the result of implementing certain constraints in the quantum theory, which classically correspond to gauge conditions fixing the densitized triad to be diagonal \cite{Makinen:2023shj}. As the formulation of the model takes place within the kinematical framework of full loop quantum gravity, quantum-reduced loop gravity enjoys a relatively clear and transparent relation with the full theory \cite{Makinen:2023shj, Makinen:2020rda, Alesci:2013xya} in comparison with loop quantum cosmology and related models, which are based on a reduction of the degrees of freedom already at the level of the classical theory.

From the point of view of practical applications, an essential advantage of quantum-reduced loop gravity is the remarkable simplicity of the model's basic kinematical structures, namely the states spanning the Hilbert space of the model and the fundamental operators acting thereon, when compared with their counterparts in full loop quantum gravity. A notable example is the volume operator, which, apart from being an important geometrical observable, is also relevant as a key building block entering the construction of any Hamiltonian constraint operator typically considered in loop quantum gravity. In quantum-reduced loop gravity, the action of the volume operator is diagonal on the natural basis states of the model, in spite of the fact that the matrix elements of the volume operator in the spin network basis of full loop quantum gravity cannot be evaluated in explicit analytic form, except in certain simple special cases. By virtue of its simplified kinematical structure, quantum-reduced loop gravity enables one to perform concrete calculations with relative ease, even when the corresponding calculation in the setting of full loop quantum gravity may be extremely challenging, if not completely intractable.

In this article we construct a concrete implementation of a Hamiltonian constraint operator for quantum-reduced loop gravity, with a view towards examining this operator in the context of a simple example, in which we consider states whose graph consists of a single six-valent node formed by three orthogonal edges (the edges being embedded in a spatial manifold having the topology of a three-torus, or carrying periodic boundary conditions). The Euclidean part of the Hamiltonian constraint is obtained through a simple modification of a construction considered earlier in \cite{Alesci:2015wla, Yang:2015zda, Assanioussi:2017tql}, while the Lorentzian part is represented by the scalar curvature operator introduced in \cite{Lewandowski:2021iun, Lewandowski:2022xox}. We derive explicit expressions representing the action of the Hamiltonian on the reduced spin network basis. Applying these general expressions to the example of the one-vertex states described above reveals a certain formal similarity between the Euclidean part of the Hamiltonian acting on the one-vertex states, and the Hamiltonian constraint of models describing an anisotropic Bianchi I universe in loop quantum cosmology.

Carrying over this formal analogy to the Lorentzian part of the Hamiltonian, we are led to propose a possible alternative treatment of the Hamiltonian constraint in loop quantum cosmology, in which the Lorentzian term (corresponding to the scalar curvature of the spatial manifold) is not identically vanishing, but is represented by a non-trivial operator which merely reduces to zero in the limit where the ``polymerization parameter'' $\mu$ is taken to zero. We emphasize that for now this modified Hamiltonian should be considered as a purely heuristic, even if potentially intriguing proposal, being motivated merely by the similarity in structure between Bianchi I loop quantum cosmology and the model of one-vertex states in quantum-reduced loop gravity. Establishing a systematic derivation of the new Lorentzian term, for instance along the lines of \cite{Dapor:2017rwv, Assanioussi:2019iye}, where a different proposal for the Lorentzian term in loop quantum cosmology is derived by considering the expectation value of a Hamiltonian constraint operator in loop quantum gravity on semiclassical states peaked on homogeneous and isotropic spatial geometries, is left as a question to be addressed in future work.

The relation between quantum-reduced loop gravity and loop quantum cosmology has previously been examined in several articles (for a non-exhaustive selection, see \eg \cite{Alesci:2013xd, Alesci:2014rra, Alesci:2015nja, Alesci:2016gub}). A common feature shared among all of these works is that the Lorentzian part of the Hamiltonian constraint operator is assumed to vanish, due to the fact that the spatial curvature of a homogeneous and isotropic, or Bianchi I universe is classically vanishing. Thus a key aspect, which distinguishes the work presented in this article from previous related work, is that here the scalar curvature is represented in the quantum theory by a non-trivial operator, which originates from an operator defined in the framework of full loop quantum gravity (although restricted to cubical graphs) \cite{Lewandowski:2021iun, Lewandowski:2022xox}. Within the covariant (spin foam) formulation of loop quantum gravity dynamics, transition amplitudes between simple quantum states describing cosmological geometries have been studied \eg in \cite{Bianchi:2010zs, Vidotto:2011qa, Bianchi:2011ym, Rennert:2013pfa, Rennert:2013qsa}; in particular, in \cite{Rennert:2013pfa, Rennert:2013qsa} states defined on a graph formed by a single six-valent vertex were considered. At this stage we cannot make any definite statement on the relation between these works and the results obtained in the present article, since in this article no concrete examples of the dynamics are analyzed; however, a more direct comparison may become possible in the future, when the dynamics of specific quantum states are evaluated in the one-vertex model introduced in section \ref{sec:1v-model}.

The material in this article is organized as follows. The present introductory section is followed by section \ref{sec:QRLG}, in which we give a brief overview of the kinematical framework of quantum-reduced loop gravity. Then, in section \ref{sec:hamiltonian} we formulate a concrete definition of a Hamiltonian constraint operator for quantum-reduced loop gravity and provide expressions representing the action of the Hamiltonian on generic reduced spin network states. In section \ref{sec:one-vertex}, these general expressions are applied to the example of reduced spin network states consisting only of a single vertex. We argue that the formal similarity between the Euclidean part of the Hamiltonian in the one-vertex example and the Hamiltonian constraint of Bianchi I models of loop quantum cosmology suggests the possibility of a modified formulation of the Hamiltonian constraint in loop quantum cosmology. The results obtained in this article are summarized in the concluding section \ref{sec:conclusions}. The article includes an appendix in which we derive some useful relations involving the eigenstates of the angular momentum operator.

\section{Quantum-reduced loop gravity}
\label{sec:QRLG}

\subsection{The reduced Hilbert space}

The Hilbert space of quantum-reduced loop gravity is a specific subspace of the kinematical Hilbert space of loop quantum gravity. It is obtained as the result of a quantum gauge-fixing procedure, where a set of gauge conditions classically fixing the densitized triad to be diagonal are implemented in the quantum theory (e.g. by means of a master constraint operator \cite{Makinen:2023shj}) in order to select a sector of the full kinematical Hilbert space as the Hilbert space of the quantum-reduced model. A basis on this Hilbert space is formed by the so-called reduced spin network states. From the viewpoint of full loop quantum gravity, a reduced spin network state is a (generalized, non-gauge invariant) spin network state, which is characterized by the following requirements:
\begin{itemize}
	\item The graph on which the state is defined is cubical, \ie the nodes (vertices) of the graph are six-valent, and the edges are aligned with the coordinate directions corresponding to a fixed Cartesian background coordinate system. For concreteness, we assume that the orientation of each edge of the graph agrees with the positive direction of the corresponding background coordinate axis.
	\item The spin quantum number associated to each edge is large, \ie
		\begin{equation}
			j_e \gg 1
			\label{}
		\end{equation}
		for every edge $e$ of the graph.
	\item The $SU(2)$ representation matrix associated to each edge is labeled by magnetic quantum numbers taking the maximal\footnote{
			In complete generality, the Hilbert space of quantum-reduced loop gravity also includes states for which some of the magnetic quantum numbers take the minimal value ($m_e = -j_e$), with certain matching conditions satisfied by the magnetic numbers at each node of the graph \cite{Makinen:2023shj}. However, in this article we consider exclusively the sector spanned by the states in which all magnetic numbers are maximal.
		} value (\ie $m_e = j_e$) with respect to the basis which diagonalizes the angular momentum component corresponding to the direction of the edge. (The precise meaning of this statement is expressed by \Eqs \eqref{eq:Dmn_i} and \eqref{eq:basis} below.)
\end{itemize}
In order to write down the basis states spanning the reduced Hilbert space, let us introduce some helpful notation. We denote by
\begin{equation}
	\ket{jm}_i \qquad (i = x, y, z)
	\label{}
\end{equation}
the eigenstate\footnote{The states $\ket{jm}_i$ can be constructed by starting with the standard basis $\ket{jm}$, in which $J_z$ is diagonal, and applying an $SU(2)$ transformation which rotates the $z$-axis into the $i$-axis:
	\[
		\ket{jm}_i = D^{(j)}(g_i)\ket{jm}.
	\]
	In order to uniquely fix the group element $g_i$, we require that the associated rotation corresponds to a cyclic permutation of the coordinate axes. See Appendix \ref{sec:J} for more details.
} of the angular momentum component $J_i$ (that is, $\ket{jm}_i$ is an eigenstate of $J^2$ and $J_i$ with respective eigenvalues $j(j+1)$ and $m$), and by
\begin{equation}
	D^{(j)}_{mn}(h)_i = {}_i\bra{jm}D^{(j)}(h)\ket{jn}_i
	\label{eq:Dmn_i}
\end{equation}
the matrix elements of the $SU(2)$ representation matrices in the basis $\ket{jm}_i$. We also use the notation
\begin{equation}
	\D^{(j)}_{mn}(h)_i = \sqrt{2j+1}D^{(j)}_{mn}(h)_i
	\label{}
\end{equation}
to denote the normalized representation matrix elements (with respect to the norm defined by the $SU(2)$ Haar measure). 

Then, considering a fixed but arbitrary cubical graph $\Gamma$, the reduced Hilbert space associated with the graph is spanned by the basis states
\begin{equation}
	\prod_{e\in\Gamma} \D^{(j_e)}_{j_ej_e}(h_e)_{i_e}
	\label{eq:basis}
\end{equation}
where it is assumed that for each edge $e$, the label $i_e$ takes the value $x$, $y$ or $z$ corresponding to the direction of the edge. The states \eqref{eq:basis} are sometimes referred to as reduced spin network states.

\subsection{Reduced operators}

From the point of view of practical applications, a key advantage of quantum-reduced loop gravity is the remarkably simple form of the model's basic operators, when compared with the corresponding operators of full loop quantum gravity. In spite of their simplicity, the operators of the quantum-reduced model arise directly from the action of the operators of the full theory on the reduced Hilbert space, after the result is truncated at leading order in the spin quantum numbers \cite{Makinen:2020rda}. For most operators of practical interest, the action of a loop quantum gravity operator $\hat{\cal O}$ on the space spanned by the states \eqref{eq:Dmn_i} yields a result of the form
\begin{equation}
	\hat{\cal O}\ket{\Psi_0} = f(j)\ket{\Phi_0} + g(j)\ket{\varphi},
	\label{eq:O-full}
\end{equation}
where $\ket{\Psi_0}$ and $\ket{\Phi_0}$ are normalized states in the reduced Hilbert space, $\ket{\varphi}$ is a normalized state generally not belonging to the reduced Hilbert space, and as long as the spin quantum numbers are assumed to be large, the first term on the right-hand side dominates over the second one:
\begin{equation}
	f(j) \gg g(j).
	\label{eq:f>>g}
\end{equation}
From any operator $\hat{\cal O}$ whose action on reduced spin network states has the form \eqref{eq:O-full}, one can obtain a corresponding reduced operator ${}^R\hat{\cal O}$ by neglecting the second term on the right-hand side of \Eq \eqref{eq:O-full}. (Due to \Eq \eqref{eq:f>>g}, the discarded term is guaranteed to be small relative to the term being kept.) Thus, the action of the reduced operator on states in the reduced Hilbert space is defined to be
\begin{equation}
	{}^R\hat{\cal O}\ket{\Psi_0} = f(j)\ket{\Phi_0}.
	\label{eq:O-reduced}
\end{equation}
As far as quantum-reduced loop gravity is concerned, the reduced operator ${}^R\hat{\cal O}$ is a well-defined operator, whose action preserves the Hilbert space of the model. On the other hand, as shown by
\begin{equation}
	\frac{\bigl|\bigl|\hat{\cal O}\ket{\Psi_0} - {}^R\hat{\cal O}\ket{\Psi_0}\bigr|\bigr|}{\bigl|\bigl|\hat{\cal O}\ket{\Psi_0}\bigr|\bigr|} \ll 1,
	\label{}
\end{equation}
from the point of view of full loop quantum gravity the reduced operator ${}^R\hat{\cal O}$ can be viewed as an accurate approximation of the action of the full operator $\hat{\cal O}$ on reduced spin network states.

As a simple illustration, let us compare the elementary operators of loop quantum gravity with their reduced counterparts. This example also serves to complete our overview of the kinematical structure of quantum-reduced loop gravity by introducing the elementary operators of the model. The holonomy operator of full loop quantum gravity acts on the spin network basis as a multiplicative operator. Its action on the basis states $D^{(j)}_{mn}(h_e)$ is given by the $SU(2)$ Clebsch--Gordan series as
\begin{equation}
	\widehat{D^{(s)}_{mn}(h_e)} D^{(j)}_{m'n'}(h_e) = \sum_k C^{(s\;j\;k)}_{m\;m'\;m+m'}C^{(s\;j\;k)}_{n\;n'\;n+n'}D^{(k)}_{m+m'\;n+n'}(h_e)
	\label{eq:D-full}
\end{equation}
where $C^{(j_1\;j_2\;j)}_{m_1\;m_2\;m}$ are the Clebsch--Gordan coefficients of $SU(2)$. As an operator conjugate to the holonomy, we consider the spin operator $\hat J_i^{(v, e)}$. This operator is labeled by a point $v$ and an edge $e$, and it carries an $su(2)$ algebra index $i$. It acts on the basis state $D^{(j)}_{mn}(h_e)$ as the left- or right-invariant vector field of $SU(2)$, according to whether the point $v$ is the beginning or ending point of the edge $e$:
\begin{equation}
	\hat J_i^{(v, e)}D^{(j)}_{mn}(h_e) = \begin{cases}
		iD^{(j)}_{mm'}(h_e)\bigl(\tau^{(j)}_i\bigr)_{m'n} & \text{$e$ begins from $v$} \\[1ex]
		-i\bigl(\tau^{(j)}_i\bigr)_{mm'}D^{(j)}(h_e)_{m'n} & \text{$e$ ends at $v$}
	\end{cases}
	\label{eq:J-full}
\end{equation}
Here $\tau^{(j)}_i$ are the anti-Hermitian generators of $SU(2)$ in the spin-$j$ representation.

The elementary operators of quantum-reduced loop gravity can be obtained by applying \Eqs \eqref{eq:D-full} and \eqref{eq:J-full} to the basis states \eqref{eq:basis} and extracting the terms of leading order in $j$ in the resulting expressions \cite{Makinen:2020rda}. In this way one finds that the reduced holonomy and spin operators act on the reduced basis states as
\begin{align}
	{}^R\widehat{D^{(s)}_{mn}(h_e)_{i_e}}\D^{(j)}_{jj}(h_e)_{i_e} &= \delta^m_n\D^{(j+m)}_{j+m\; j+m}(h_e)_{i_e} \label{eq:D-reduced} \\[2ex]
	{}^R\hat J_i^{(v, e)}\D^{(j)}_{jj}(h_e)_{i_e} &= \pm\delta_i^{i_e}j\D^{(j)}_{jj}(h_e)_{i_e}
	\label{eq:J-reduced}
\end{align}
In \Eq \eqref{eq:D-reduced} it is assumed that $s \ll j$, \ie the spin carried by the holonomy operator is small compared to the spin quantum number of the state on which the operator is acting, while in \Eq \eqref{eq:J-reduced} the plus and minus signs correspond, as in \Eq \eqref{eq:J-full}, to $v$ being respectively the beginning point and endpoint of the edge $e$. It is also worth noting that the multiplication law \eqref{eq:D-reduced} for reduced holonomies holds only if the holonomies are expressed in the correct basis, \ie the eigenbasis of the angular momentum component corresponding to the direction of the edge $e$.

Comparing the structure of \Eqs \eqref{eq:D-reduced} and \eqref{eq:J-reduced} with \Eqs \eqref{eq:D-full} and \eqref{eq:J-full}, we can see that the simplification achieved by the quantum-reduced model is twofold. Firstly, only certain components of the elementary reduced operators have a non-vanishing action on the reduced Hilbert space. These are the diagonal $(m=n)$ components of the reduced holonomy operator, and the component $i=x$, $y$ or $z$ of the reduced spin operator which matches the direction of the edge $e$. Secondly, the matrix elements of the reduced operators are considerably simpler than those of their counterparts in the full theory. The action of the reduced spin operator on the reduced basis states is simply diagonal, while the reduced holonomy operator acts essentially according to a $U(1)$ multiplication law (in which the $U(1)$ ``charge'' carried by the operator is determined by the magnetic quantum number instead of the spin).

\begin{figure}[t]
	\centering
	\includegraphics[scale=0.18]{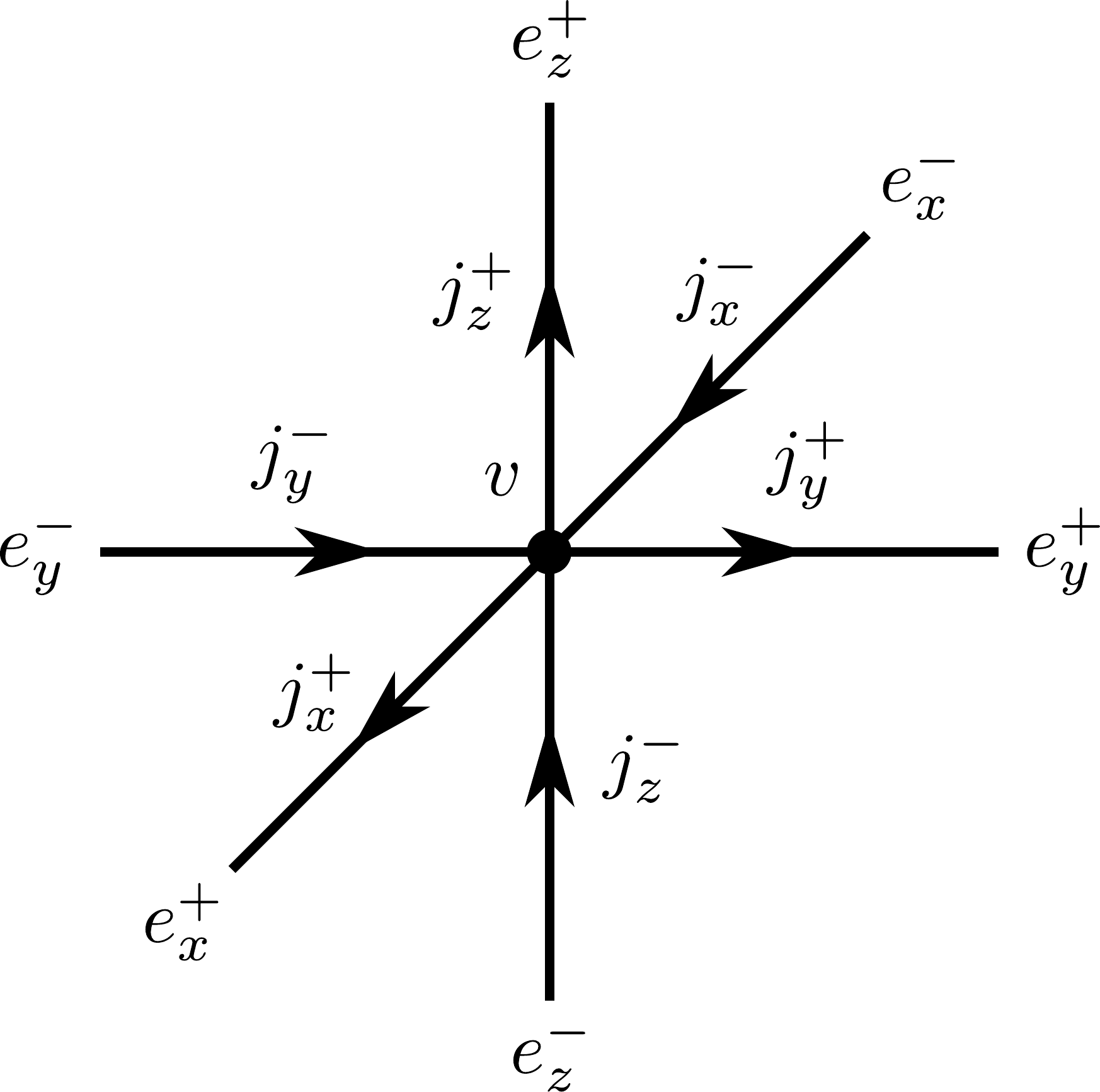}
	\caption{Labeling of the edges and spin quantum numbers at a given node $v$ of a reduced spin network state. The edge $e_a^\pm$ $(a = x, y, z)$ is aligned with the $x^a$-coordinate direction defined by a fixed Cartesian background coordinate system, and lies before ($-$) or after ($+$) the node $v$ -- in the sense of the positive direction of the $x^a$-coordinate axis -- as indicated by the superscript. (Figure reproduced from \cite{Makinen:2023shj}.)}
	\label{fig:node}
\end{figure}

Let us conclude this section by introducing the reduced volume operator, which will play an important role in the next section as an ingredient of the Hamiltonian constraint operator. The Ashtekar--Lewandowski volume operator \cite{Ashtekar:1997fb} of loop quantum gravity, restricted to a single node $v$ of an arbitrary spin network state, is given by
\begin{equation}
	\hat V_v = \sqrt{\biggl|\dfrac{1}{48}\sum_{IJK}\epsilon(e_I, e_J, e_K) \epsilon^{ijk}\hat J_i^{(v, e_I)}\hat J_j^{(v, e_J)}\hat J_k^{(v, e_K)}\biggr|}
	\label{eq:vol}
\end{equation}
where $I$, $J$ and $K$ enumerate the edges incident on the node, and the geometric factor $\epsilon(e_I, e_J, e_K) = +1, 0, -1$ is essentially the orientation of the triple of tangent vectors $(\dot e_I, \dot e_J, \dot e_K)$ at $v$. In the context of full loop quantum gravity, the volume operator is a rather complicated object, whose action on generic spin network states cannot be evaluated in explicit analytic form. Nevertheless, by examining the action of the volume operator on the states \eqref{eq:basis} it can be shown that the reduced volume operator emerging from \Eq \eqref{eq:vol} is diagonal on the reduced spin network basis \cite{Makinen:2020rda}. Letting $\ket{\Psi_0}$ denote a basis state of the form \eqref{eq:basis}, we have
\begin{equation}
	{}^R\hat V_v\ket{\Psi_0} = \Upsilon_v\ket{\Psi_0}
	\label{}
\end{equation}
where the eigenvalue is
\begin{equation}
	\Upsilon_v = \sqrt{\frac{1}{8}\bigl(j_x^+ + j_x^-\bigr)\bigl(j_y^+ + j_y^-\bigr)\bigl(j_z^+ + j_z^-\bigr)}
	\label{}
\end{equation}
and $j_i^\pm$ denote the spins on the edges at the node $v$ as shown in \Fig \ref{fig:node}.

\section{The Hamiltonian constraint}
\label{sec:hamiltonian}

In the canonical formulation of loop quantum gravity, the dynamics of the theory is determined by the Hamiltonian constraint operator (see \eg \cite{Thiemann:1996aw, Lewandowski:2014hza, Assanioussi:2015gka, Yang:2015zda}). The role of the Hamiltonian constraint in the quantum theory is to determine the physical Hilbert space and the physical observables of the theory. In the present work we take the classical Hamiltonian constraint to be represented by the functional
\begin{equation}
	C(N) = \frac{1}{\beta^2}\int d^3x\, N\biggl(\frac{\epsilon\updown{ij}{k}E^a_iE^b_jF_{ab}^k}{\sqrt{|\det E|}} + (1 + \beta^2)\sqrt{|\det E|}\Rt\biggr)
	\label{eq:C}
\end{equation}
where $\beta$ is the Barbero--Immirzi parameter, $N$ is the lapse function, $E^a_i$ is the densitized triad, $F_{ab}^i$ is the curvature of the Ashtekar--Barbero connection, and the Lorentzian part of the Hamiltonian constraint is represented by the term involving the Ricci scalar $\Rt$ of the spatial manifold.

As an alternative to the fully constrained formulation of the dynamics in vacuum loop quantum gravity, one may consider the deparametrized approach, in which a scalar matter field $\phi$ is employed as a relational time variable for the dynamics of the gravitational field \cite{Rovelli:1993bm, Domagala:2010bm, Giesel:2012rb, Assanioussi:2017tql}. In deparametrized models of loop quantum gravity, the dynamics of (diffeomorphism invariant) quantum states of the gravitational field is generated by a physical Hamiltonian operator $\hat H_{\rm phys}$ according to the Schrödinger equation
\begin{equation}
	i\frac{d}{d\phi}\ket{\Psi} = \hat H_{\rm phys}\ket{\Psi}.
	\label{}
\end{equation}
The physical Hamiltonian $\hat H_{\rm phys}$ is a quantization of a classical functional closely related to the Hamiltonian constraint, its precise form depending on the type of scalar field chosen as the physical time variable. In particular, if the reference field is taken to be an irrotational dust field \cite{Brown:1994py, Husain:2011tk, Swiezewski:2013jza}, then the physical Hamiltonian is given directly by the Hamiltonian constraint evaluated at unit lapse function:
\begin{equation}
	\hat H_{\rm phys} = \hat C(1).
	\label{eq:H_phys}
\end{equation}

Our goal in this section is to provide a concrete example of a Hamiltonian constraint operator for quantum-reduced loop gravity. According to the general scheme expressed by \Eqs \eqref{eq:O-full}--\eqref{eq:O-reduced}, this is achieved by starting with an operator defined in the framework of full loop quantum gravity and identifying the terms of leading order in $j$ in the action of the operator on the reduced basis states \eqref{eq:basis} in order to establish the corresponding reduced operator. In section \ref{sec:C_cubic} we summarize the definition of the Hamiltonian constraint operator which will be considered in this article. This definition is valid not only in the setting of quantum-reduced loop gravity but applies more generally to all states based on cubical graphs in the kinematical Hilbert space of loop quantum gravity. The Euclidean part of the Hamiltonian will be represented by an operator which is known in the literature of loop quantum gravity \cite{Alesci:2015wla, Yang:2015zda, Assanioussi:2017tql} but has not been previously applied to quantum-reduced loop gravity. In section \ref{sec:C_E} we will therefore present in detail the calculations required to obtain the reduced operator corresponding to this definition of the Euclidean part of the Hamiltonian. For the Lorentzian part of the Hamiltonian, we will use the scalar curvature operator introduced in \cite{Lewandowski:2021iun} and studied in the context of quantum-reduced loop gravity in \cite{Lewandowski:2022xox}. For the sake of completeness, in section \ref{sec:C_L} we will recap the result established in \cite{Lewandowski:2022xox} for the curvature operator on the reduced Hilbert space.

\subsection{Hamiltonian constraint operator on a cubical graph}
\label{sec:C_cubic}

For the Euclidean part of the Hamiltonian constraint, namely the operator corresponding to the classical functional
\begin{equation}
	C_E(N) = \int d^3x\, N\frac{\epsilon\updown{ij}{k}E^a_iE^b_jF_{ab}^k}{\sqrt{|\det E|}}
	\label{}
\end{equation}
we adopt a modified version of the construction considered previously in \cite{Alesci:2015wla, Assanioussi:2017tql} in the context of deparametrized models of loop quantum gravity, and in \cite{Yang:2015zda} as a constraint operator for the vacuum theory. On a given graph $\Gamma$, the Euclidean Hamiltonian will be represented by the operator
\begin{equation}
	\hat C_E(N) = \sum_{v\in\Gamma} N(v) \sum_{\text{$e{\nparallel}e'$ at $v$}} \hat C_E^{(v, e, e')}
	\label{eq:C_E}
\end{equation}
where the inner sum runs over all pairs of edges whose tangent vectors at $v$ are linearly independent, and the operator associated to a node $v$ and a pair of edges $(e, e')$ at $v$ is given by
\begin{equation}
	\hat C_E^{(v, e, e')} = -\frac{1}{2}\epsilon^{ijk}\Tr\bigl(\tau_k \widehat{h_{\alpha_{ee'}}}\bigr)\hat J_i^{(v,e)}\hat J_j^{(v,e')}\widehat{{\cal V}_v^{-1}}.
	\label{eq:C_E^v}
\end{equation}
Here
\begin{equation}
	\widehat{{\cal V}_v^{-1}} = \lim_{\epsilon\to 0} \frac{\hat V_v}{\hat V_v^2 + \epsilon^2}
	\label{eq:V^-1}
\end{equation}
is the so-called Tikhonov regularization of the inverse volume operator\footnote{
	Equivalently, given a complete set of eigenstates $\ket{\lambda}$ of the volume operator $\hat V_v$ with corresponding eigenvalues $\lambda$, the operator $\widehat{{\cal V}_v^{-1}}$ can be defined by
	\[
		\widehat{{\cal V}_v^{-1}}\ket{\lambda} = \begin{cases}
			\lambda^{-1}\ket{\lambda} & \text{if $\lambda\neq 0$} \\
			0 & \text{if $\lambda = 0$}
		\end{cases}
	\]
}, and $\alpha_{ee'}$ denotes a closed loop associated with the pair $(e, e')$. In the setting of quantum-reduced loop gravity and the cubical graphs considered in the model, we choose a graph-preserving loop assignment in which the loop $\alpha_{ee'}$ is taken to be the minimal rectangular loop formed by the edges $e$ and $e'$ together with two adjacent edges of the cubical graph. We further assume that the trace in \Eq \eqref{eq:C_E^v} is taken in the fundamental representation.

For the Lorentzian part of the Hamiltonian constraint, we follow the quantization presented in \cite{Lewandowski:2021iun, Lewandowski:2022xox}, where an operator representing the integrated Ricci scalar $\int d^3x\,\sqrt q\Rt$ is first constructed for arbitrary states defined on cubical graphs in the Hilbert space of loop quantum gravity, and then an operator for quantum-reduced loop gravity is derived from the more general operator by applying this operator on the basis states \eqref{eq:basis} and extracting the terms of leading order in $j$. The starting point of the construction performed in \cite{Lewandowski:2021iun} is to write the Lorentzian part of the classical Hamiltonian constraint in the form
\begin{equation}
	C_L(N) = \int d^3x\, N\sqrt{|\det E|}\Rt\bigl[E^a_i, {\cal D}_aE^b_i, {\cal D}_a{\cal D}_bE^c_i\bigr]
	\label{eq:C_L}
\end{equation}
where the Ricci scalar is expressed as a function of the densitized triad and its gauge covariant derivatives (\ie ${\cal D}_a E^b_i = \partial_a E^b_i + \epsilon\downup{ij}{k}A_a^jE^b_k$, where $A_a^i$ is the Ashtekar connection). Given a cubical graph, the covariant derivatives can then be quantized by regularizing them in terms of finite differences of parallel transported flux variables (also known in the literature as gauge covariant fluxes) on the cubic lattice provided by the graph. As a result one obtains an operator which represents a quantization of the classical functional \eqref{eq:C_L} on the kinematical Hilbert space of loop quantum gravity restricted to a fixed but arbitrary cubical graph. This operator has the form
\begin{equation}
	\hat C_L(N) = \sum_{v\in\Gamma} N(v)\hat{\cal R}_v\widehat{{\cal V}_v^{-1}}
	\label{eq:C_L^}
\end{equation}
where $\hat{\cal R}_v$ is an operator which can be expressed as a combination of the operators $\hat J_i^{(v,e)}$ defined by \Eq \eqref{eq:J-full}, and holonomy operators associated with the edges of the cubical graph $\Gamma$. (For an explicit definition of $\hat{\cal R}_v$, we refer the reader to \cite{Lewandowski:2021iun}. Note also that in \Eq \eqref{eq:C_L^} we have chosen a reversed factor ordering between the operators $\hat{\cal R}_v$ and $\widehat{{\cal V}_v^{-1}}$ compared to the choice originally made in \cite{Lewandowski:2021iun}.)

The operators representing the Euclidean and Lorentzian parts of the Hamiltonian having been defined, the complete Hamiltonian constraint operator now is
\begin{equation}
	\hat C(N) = \frac{1}{\beta^2}\hat C_E(N) + \frac{1 + \beta^2}{\beta^2}\hat C_L(N)
	\label{eq:C^}
\end{equation}
where $\hat C_E(N)$ and $\hat C_L(N)$ are given respectively by \Eqs \eqref{eq:C_E} and \eqref{eq:C_L^}. The definition \eqref{eq:C^} is valid for all states based on cubical graphs in the kinematical Hilbert space of loop quantum gravity. A Hamiltonian constraint operator for quantum-reduced loop gravity can then be obtained by applying the operator \eqref{eq:C^} on the Hilbert space of the quantum-reduced model and identifying the contribution of leading order in the spin quantum numbers as the corresponding reduced operator.

\subsection{Reduced operator: Euclidean part}
\label{sec:C_E}

\begin{figure}[t]
	\centering
	\includegraphics[scale=0.18]{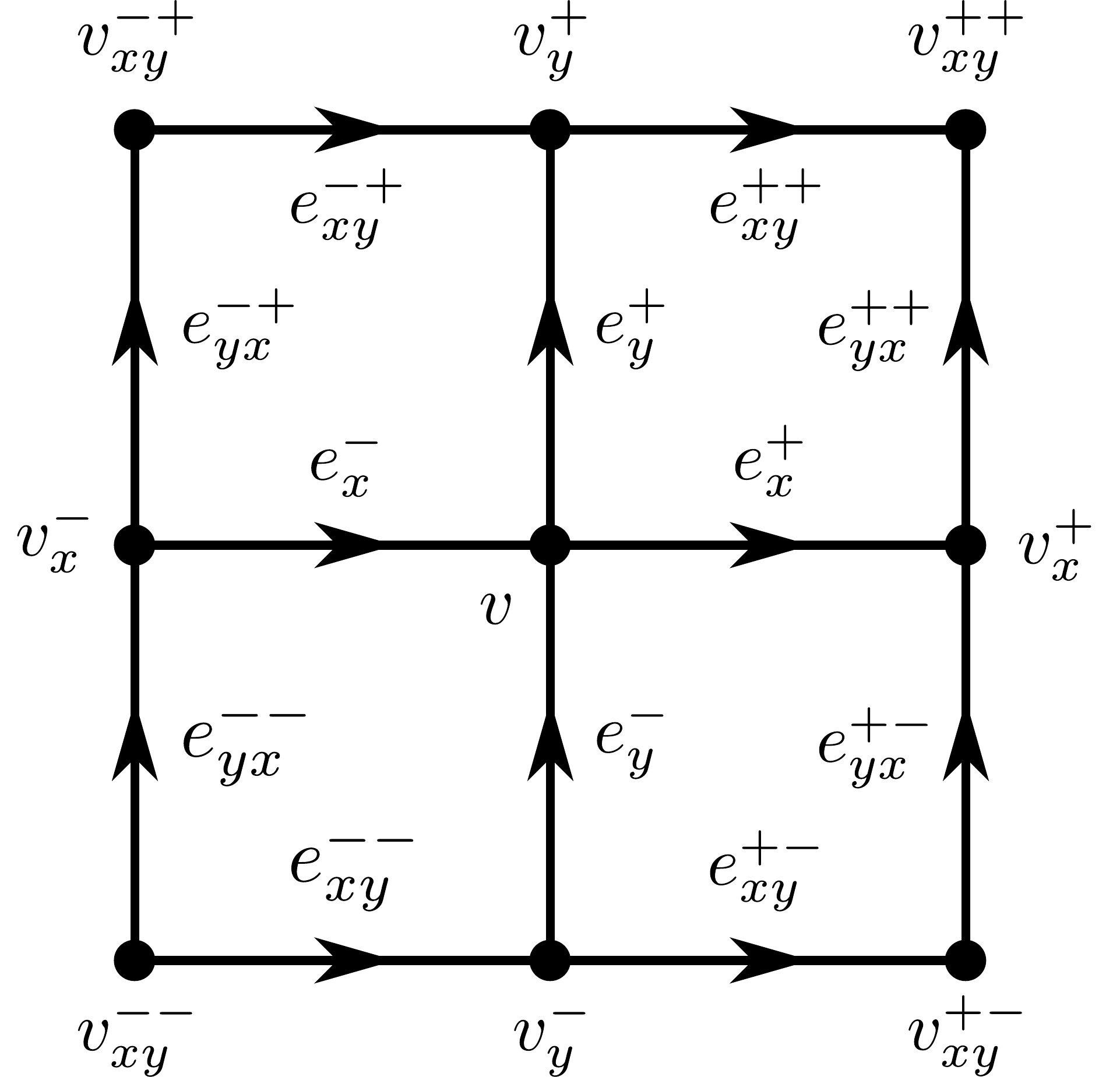}
	\caption{Labeling of the edges and nodes surrounding a given node $v$ of a cubical graph. (A similar labeling applies to edges and nodes lying in the $xz$- and $yz$-coordinate planes.) For the edges and nodes not immediately adjacent to $v$, the subscript denotes the coordinate plane and the superscript the quadrant in which the node or edge lies. Moreover, for the edges the first index of the superscript indicates the direction of the edge; for example, the edges $e_{xy}^{++}$ and $e_{yx}^{++}$ both lie in the first quadrant of the $xy$-plane, and are oriented respectively along the $x$- and $y$-coordinate directions.}
	\label{fig:labels}
\end{figure}

In order to derive the form of the reduced operator corresponding to the Euclidean Hamiltonian defined by \Eqs \eqref{eq:C_E} and \eqref{eq:C_E^v}, let us first focus on a single pair of edges at $v$, for instance $(e_x^+, e_y^+)$ according to the labeling displayed in \Fig \ref{fig:labels}. Then we have to consider the operator
\begin{equation}
	\hat C_E^{(v, e_x^+, e_y^+)} = -\frac{1}{2}\epsilon^{ijk}\Tr\Bigl(\tau_k h_{e_y^+}^{-1} h_{e_{xy}^{++}}^{-1} h_{e_{yx}^{++}} h_{e_x^+}\Bigr)\hat J_i^{(v,e_x^+)}\hat J_j^{(v,e_y^+)}\widehat{{\cal V}_v^{-1}},
	\label{eq:Cxy++}
\end{equation}
where we have decomposed the loop holonomy into a product of the four edge holonomies which constitute the loop. For the factor on the right of the trace, it is immediate to see that its contribution to the reduced operator is
\begin{equation}
	{}^R\Bigl(\hat J_i^{(v,e_x^+)}\hat J_j^{(v,e_y^+)}\widehat{{\cal V}_v^{-1}}\Bigr)\ket{\Psi_0} = \delta_i^x\delta_j^y\frac{j_x^+j_y^+}{\Upsilon_v}\ket{\Psi_0},
	\label{eq:right_R}
\end{equation}
with $\ket{\Psi_0}$ denoting a generic reduced basis state of the form \eqref{eq:basis}. Hence it remains to extract the contribution coming from the trace
\begin{equation}
	\Tr\Bigl(\tau_z h_{e_y^+}^{-1} h_{e_{xy}^{++}}^{-1} h_{e_{yx}^{++}} h_{e_x^+}\Bigr).
	\label{eq:Tr}
\end{equation}
Recall from \Eq \eqref{eq:D-reduced} that the reduced holonomy operator associated with a given edge is formed by the diagonal matrix elements of the holonomy with respect to the eigenbasis of the angular momentum component $J_i$ corresponding to the direction of the edge. Therefore the reduced operator corresponding to the trace \eqref{eq:Tr} can be obtained by expanding the trace, say in the eigenbasis of $J_z$, as
\begin{equation}
	(\tau_z)\updown{A}{B}\bigl(h_{e_y^+}^{-1}\bigr)\updown{B}{C}\bigl(h_{e_{xy}^{++}}^{-1}\bigr)\updown{C}{D}(h_{e_{yx}^{++}})\updown{D}{E}(h_{e_x^+})\updown{E}{A},
	\label{eq:Tr-expanded}
\end{equation}
transforming the matrix elements of each holonomy $h_e$ to the eigenbasis dictated by the direction of the edge $e$, and then discarding all terms except those involving only diagonal matrix elements.

Before proceeding to analyze the trace \eqref{eq:Tr}, it may be helpful to consider the simpler example of computing the action of a specific matrix element of the reduced holonomy operator on the basis states \eqref{eq:basis}. Let us therefore look at extracting the reduced operator corresponding to the matrix element
\begin{equation}
	(h_{e_x})\updown{+}{+} = \bra{+}h_{e_x}\ket{+}
	\label{eq:hx++}
\end{equation}
where $e_x$ is an edge oriented in the $x$-direction. The expression of the state $\ket{+}$ in the basis $\ket{\pm}_x$ is given by \Eq \eqref{eq:+(x)} as
\begin{equation}
	\ket{+} = \frac{1}{\sqrt 2}e^{i\pi/4}\ket{+}_x - \frac{1}{\sqrt 2}e^{-i\pi/4}\ket{-}_x.
	\label{}
\end{equation}
Using this in \Eq \eqref{eq:hx++}, we find
\begin{equation}
	(h_{e_x})\updown{+}{+} = \frac{1}{2}\Bigl({}_x\bra{+}h_{e_x}\ket{+}_x + {}_x\bra{-}h_{e_x}\ket{-}_x\Bigr) + \frac{i}{2}\Bigl({}_x\bra{+}h_{e_x}\ket{-}_x - {}_x\bra{-}h_{e_x}\ket{+}_x\Bigr).
	\label{}
\end{equation}
Now the reduced operator is formed by the terms containing diagonal matrix elements with respect to the $x$-basis, while the off-diagonal matrix elements do not contribute to the reduced operator. As the result of the example, we have therefore established
\begin{equation}
	{}^R(h_{e_x})\updown{+}{+} = \frac{1}{2}\Bigl({}^R\widehat{D^{(1/2)}_{++}(h_{e_x})_x} + {}^R\widehat{D^{(1/2)}_{--}(h_{e_x})_x}\Bigr),
	\label{}
\end{equation}
with the action of the operators on the right-hand side on the reduced Hilbert space being given by \Eq \eqref{eq:D-reduced}, as the reduced operator corresponding to the matrix element \eqref{eq:hx++}.

In order to carry out the analogous calculation for the trace \eqref{eq:Tr}, it is helpful to first form the matrices
\begin{equation}
	h_{e_i} = \begin{pmatrix}
		\bra{+}h_{e_i}\ket{+} & \bra{+}h_{e_i}\ket{-} \\
		\bra{-}h_{e_i}\ket{+} & \bra{-}h_{e_i}\ket{-}
	\end{pmatrix}
	\label{}
\end{equation}
for $i = x$ and $y$, transform the states $\ket{\pm}$ to the basis $\ket{\pm}_i$ and discard the off-diagonal matrix elements in the resulting expressions. Using \Eqs \eqref{eq:+(x)}--\eqref{eq:-(y)}, we find
\begin{equation}
	h_{e_x} = \begin{pmatrix} c(e_x) & is(e_x) \\ is(e_x) & c(e_x) \end{pmatrix} + \text{off-diag.}
	\label{eq:hx_R}
\end{equation}
and
\begin{equation}
	h_{e_y} = \begin{pmatrix} c(e_y) & s(e_y) \\ -s(e_y) & c(e_y) \end{pmatrix} + \text{off-diag.}
	\label{eq:hy_R}
\end{equation}
where we have introduced the abbreviations
\begin{align}
	c(e) &= \frac{1}{2}\Bigl({}_{i_e}\bra{+}h_e\ket{+}_{i_e} + {}_{i_e}\bra{-}h_e\ket{-}_{i_e}\Bigr), \notag \\
	s(e) &= \frac{1}{2i}\Bigl({}_{i_e}\bra{+}h_e\ket{+}_{i_e} - {}_{i_e}\bra{-}h_e\ket{-}_{i_e}\Bigr).
	\label{}
\end{align}
Then we can proceed directly by inserting \Eqs \eqref{eq:hx_R} and \eqref{eq:hy_R} into \Eq \eqref{eq:Tr} and evaluating the trace. This yields
\begin{align}
	&\Tr\Bigl(\tau_z h_{e_y^+}^{-1} h_{e_{xy}^{++}}^{-1} h_{e_{yx}^{++}} h_{e_x^+}\Bigr) \notag \\
	&= c(e_y^+)c(e_{xy}^{++})s(e_{yx}^{++})s(e_x^+) + c(e_y^+)s(e_{xy}^{++})s(e_{yx}^{++})c(e_x^+) \notag \\
	&\quad + s(e_y^+)s(e_{xy}^{++})c(e_{yx}^{++})c(e_x^+) - s(e_y^+)c(e_{xy}^{++})c(e_{yx}^{++})s(e_x^+) + \text{off-diag.}
	\label{eq:Tr_R}
\end{align}
from which the reduced operator corresponding to the trace can be immediately read off simply by dropping the terms denoted by ``off-diag.''

Combining \Eqs \eqref{eq:right_R} and \eqref{eq:Tr_R} with \Eq \eqref{eq:Cxy++}, we see that the reduced operator ${}^R\hat C_E^{(v, e_x^+, e_y^+)}$ is given by
\begin{align}
	{}^R\hat C_E^{(v, e_x^+, e_y^+)}\ket{\Psi_0} = -\frac{1}{2}\frac{j_x^+ j_y^+}{\Upsilon_v}\Bigl(
		&\hat c(e_y^+)\hat c(e_{xy}^{++})\hat s(e_{yx}^{++})\hat s(e_x^+) \notag \\
		&+ \hat c(e_y^+)\hat s(e_{xy}^{++})\hat s(e_{yx}^{++})\hat c(e_x^+) \notag \\
		&+ \hat s(e_y^+)\hat s(e_{xy}^{++})\hat c(e_{yx}^{++})\hat c(e_x^+) \notag \\
		&- \hat s(e_y^+)\hat c(e_{xy}^{++})\hat c(e_{yx}^{++})\hat s(e_x^+) \Bigr)\ket{\Psi_0}
	\label{eq:Cxy++_R}
\end{align}
where
\begin{align}
	\hat c(e) &= \frac{1}{2}\Bigl({}^R\widehat{D^{(1/2)}_{++}(h_e)_{i_e}} + {}^R\widehat{D^{(1/2)}_{--}(h_e)_{i_e}}\Bigr),
	\label{eq:c(e)} \\
	\hat s(e) &= \frac{1}{2i}\Bigl({}^R\widehat{D^{(1/2)}_{++}(h_e)_{i_e}} - {}^R\widehat{D^{(1/2)}_{--}(h_e)_{i_e}}\Bigr).
	\label{eq:s(e)}
\end{align}
Next, by repeating the calculation presented above for all pairs of edges entering the sum in \Eq \eqref{eq:C_E}, we find that the reduced operator associated with each pair is of the form \eqref{eq:Cxy++_R}. Thus we conclude that the Euclidean part of the Hamiltonian constraint operator for quantum-reduced loop gravity, emerging from the operator defined by \Eqs \eqref{eq:C_E} and \eqref{eq:C_E^v} as an operator of full loop quantum gravity, is given by
\begin{equation}
	{}^R\hat C_E(N) = \sum_{v\in\Gamma} N(v) \sum_{(a, b)}\sum_{\alpha\beta} {}^R\hat C_E^{(v, e_a^\alpha, e_b^\beta)},
	\label{eq:CE_R}
\end{equation}
where
\begin{align}
	{}^R\hat C_E^{(v, e_a^\alpha, e_b^\beta)}\ket{\Psi_0} = -\frac{1}{2}\frac{j_a^\alpha j_b^\beta}{\Upsilon_v}\Bigl(
		&\hat c(e_b^\beta)\hat c(e_{ab}^{\alpha\beta})\hat s(e_{ba}^{\alpha\beta})\hat s(e_a^\alpha) \notag \\
		&+ \hat c(e_b^\beta)\hat s(e_{ab}^{\alpha\beta})\hat s(e_{ba}^{\alpha\beta})\hat c(e_a^\alpha) \notag \\
		&+ \hat s(e_b^\beta)\hat s(e_{ab}^{\alpha\beta})\hat c(e_{ba}^{\alpha\beta})\hat c(e_a^\alpha) \notag \\
		&- \hat s(e_b^\beta)\hat c(e_{ab}^{\alpha\beta})\hat c(e_{ba}^{\alpha\beta})\hat s(e_a^\alpha) \Bigr)\ket{\Psi_0}
	\label{eq:CE^v_R}
\end{align}
and the first sum covers the pairs $(a, b) = (x, y)$, $(x, z)$ and $(y, z)$, while the labels $\alpha$ and $\beta$ independently range over the values $+$ and $-$.

\subsection{Reduced operator: Lorentzian part}
\label{sec:C_L}

We then proceed to the Lorentzian part of the Hamiltonian. The action of the operator \eqref{eq:C_L^} on the Hilbert space of quantum-reduced loop gravity gives rise to a corresponding reduced operator, which was considered in \cite{Lewandowski:2022xox}, and which can be used in the quantum-reduced model as an operator representing the Lorentzian part of the Hamiltonian constraint. It was established in \cite{Lewandowski:2022xox} that this operator has the form
\begin{equation}
	{}^R\hat C_L(N) = \sum_{v\in\Gamma} N(v) {}^R\hat{\cal R}_v{}^R\widehat{{\cal V}_v^{-1}}
	\label{eq:CL_R}
\end{equation}
where the operator ${}^R\hat{\cal R}_v$ is defined by\footnote{
	This is essentially \Eq (4.4) of \cite{Lewandowski:2022xox}, but in the present work we adopt a modified factor ordering, where all the reduced flux operators $\hat p_a(v)$ are ordered to the right.
}
\begin{align}
	{}^R\hat{\cal R}_v\ket{\Psi_0} &= \sum_a \biggl(\frac{3}{2}k_a^2\frac{\bigl(\Delta_a\Upsilon_v^2\bigr)^2}{\Upsilon_v^4} - 2k_a^2\frac{\Delta_a^2\Upsilon_v^2}{\Upsilon_v^2}\biggr)\ket{\Psi_0} \notag \\
	&+ \sum_{ab}\biggl[-2k_a\widehat{\Delta_{ab}^{(v)}E_a^b} + 2\frac{k_a^2}{k_b}\widehat{\Delta_{aa}^{(v)}E_b^b} \notag \\
	&\qquad\quad + 2\frac{\Delta_a\Upsilon_v^2}{\Upsilon_v^2} \biggl(\frac{k_a^2}{k_b}\widehat{\Delta_{a}^{(v)}E_b^b} - k_a\widehat{\Delta_{b}^{(v)}E_a^b} - k_b\widehat{\Delta_{b}^{(v)}E_b^a}\biggr)\biggr]\ket{\Psi_0} \notag \\
	&+ \sum_{abc}\threedots\biggl[-\widehat{\Delta_{a}^{(v)}E_c^a}\widehat{\Delta_{b}^{(v)}E_c^b} - \frac{1}{2}\widehat{\Delta_{a}^{(v)}E_c^b}\widehat{\Delta_{b}^{(v)}E_c^a} + \frac{5}{2}k_a^2\widehat{\Delta_{a}^{(v)}E_b^c}\widehat{\Delta_{a}^{(v)}(E^{-1})_c^b} \notag \\
		&\qquad\quad - \frac{1}{2}\frac{k_a^2}{k_b^2}\widehat{\Delta_{a}^{(v)}E_c^b}\widehat{\Delta_{a}^{(v)}E_c^b} + 2\frac{k_a}{k_b}\Bigl(\widehat{\Delta_{a}^{(v)}E_a^c}\widehat{\Delta_{c}^{(v)}E_b^b} + \widehat{\Delta_{c}^{(v)}E_a^c}\widehat{\Delta_{a}^{(v)}E_b^b}\Bigr) \notag \\
	&\qquad\quad + \frac{k_a}{k_b}\widehat{\Delta_{c}^{(v)}E_a^b}\widehat{\Delta_{a}^{(v)}E_b^c} + \frac{1}{2}\frac{k_ak_b}{k_c^2}\widehat{\Delta_{a}^{(v)}E_b^c}\widehat{\Delta_{b}^{(v)}E_a^c} - \frac{k_a^2}{k_bk_c}\widehat{\Delta_{a}^{(v)}E_b^b}\widehat{\Delta_{a}^{(v)}E_c^c}\biggr]\threedots\;\ket{\Psi_0}.
	\label{eq:R_v}
\end{align}
In \Eq \eqref{eq:R_v} the abbreviations
\begin{equation}
	k_a = \frac{1}{2}\bigl(j_a^+ + j_a^-\bigr)
	\label{}
\end{equation}
and
\begin{align}
	\Delta_a\Upsilon_v^2 &= \frac{1}{2}\Bigl(\Upsilon_{v_a^+}^2 - \Upsilon_{v_a^-}^2\Bigr) \\[1ex]
	\Delta_a^2\Upsilon_v^2 &= \Upsilon_{v_a^+}^2 - 2\Upsilon_v^2 + \Upsilon_{v_a^-}^2
	\label{}
\end{align}
are used. Moreover, the operators
\begin{equation}
	\widehat{\Delta_{a}^{(v)}E_i^b}, \qquad \widehat{\Delta_{a}^{(v)}(E^{-1})_b^i}, \qquad \widehat{\Delta_{ab}^{(v)}E_i^c}
	\label{eq:DEs}
\end{equation}
whose explicit expression will be given below, originate from the discretization of covariant derivatives of the densitized triad on the lattice provided by the cubical graph; they represent respectively the quantization of ${\cal D}_a E^b_i$, ${\cal D}_a E_b^i$ ($E_a^i$ being the inverse of $E^a_i$) and the symmetric part of the second derivative ${\cal D}_a{\cal D}_b E^c_i$. Generally speaking, the operators \eqref{eq:DEs} are formed by combinations of the reduced flux operator $\hat p_a(v)$, which is defined by its action on the basis states \eqref{eq:basis} as
\begin{equation}
	\hat p_a(v)\ket{\Psi_0} = \frac{1}{2}\Bigl(j_a^+(v) + j_a^-(v)\Bigr)\ket{\Psi_0},
	\label{}
\end{equation}
and the symmetric holonomy operators
\begin{align}
	\hat c^{(1)}(e) &= \frac{1}{2}\Bigl({}^R\widehat{D^{(1)}_{11}(h_e)_{i_e}} + {}^R\widehat{D^{(1)}_{-1\; -1}(h_e)_{i_e}} \Bigr)
	\label{eq:c(1)} \\
	\hat s^{(1)}(e) &= \frac{1}{2i}\Bigl({}^R\widehat{D^{(1)}_{11}(h_e)_{i_e}} - {}^R\widehat{D^{(1)}_{-1\; -1}(h_e)_{i_e}} \Bigr),
	\label{eq:s(1)}
\end{align}
which are the analogue of the operators \eqref{eq:c(e)} and \eqref{eq:s(e)} in the spin-1 representation. Finally, the triple dots around the last square brackets in \Eq \eqref{eq:R_v} denote a choice of factor ordering for the product of two discretized derivative operators, in which the reduced flux operator contributed by the leftmost derivative operator is ordered to the right of the reduced holonomy operators contributed by the derivative operator on the right.

For completeness, we conclude this section by giving the explicit expressions for the components of the operators \eqref{eq:DEs}. In the expressions below, the following conventions are adopted:
\begin{itemize}
	\item The indices $(a, b, c)$ are equal to any {\em cyclic} permutation of $(x, y, z)$.
	\item Only the non-vanishing components of the operators are listed; thus, any component which is not given is identically equal to zero.
\end{itemize}

\subsubsection*{The operator $\widehat{\Delta_{a}^{(v)}E_i^b}$}
\begin{align}
	\widehat{\Delta_{a}^{(v)}E_a^a} &= \frac{1}{2}\Bigl(\hat p_a(v_a^+) - \hat p_a(v_a^-)\Bigr) \label{eq:Da-first} \\
	\widehat{\Delta_{a}^{(v)}E_b^b} &= \frac{1}{2}\Bigl(\hat c^{(1)}(e_a^+)\hat p_b(v_a^+) - \hat c^{(1)}(e_a^-)\hat p_b(v_a^-)\Bigr) \\
	\widehat{\Delta_{a}^{(v)}E_c^c} &= \frac{1}{2}\Bigl(\hat c^{(1)}(e_a^+)\hat p_c(v_a^+) - \hat c^{(1)}(e_a^-)\hat p_c(v_a^-)\Bigr) \\
	\widehat{\Delta_{a}^{(v)}E_c^b} &= \frac{1}{2}\Bigl(\hat s^{(1)}(e_a^+)\hat p_b(v_a^+) + \hat s^{(1)}(e_a^-)\hat p_b(v_a^-)\Bigr) \\
	\widehat{\Delta_{a}^{(v)}E_b^c} &= -\frac{1}{2}\Bigl(\hat s^{(1)}(e_a^+)\hat p_c(v_a^+) + \hat s^{(1)}(e_a^-)\hat p_c(v_a^-)\Bigr)
	\label{}
\end{align}

\subsubsection*{The operator $\widehat{\Delta_{a}^{(v)}(E^{-1})_b^i}$}
\begin{align}
	\widehat{\Delta_{a}^{(v)}(E^{-1})_a^a} &= \frac{1}{2}\Bigl(\hat p_a^{-1}(v_a^+) - \hat p_a^{-1}(v_a^-)\Bigr) \\
	\widehat{\Delta_{a}^{(v)}(E^{-1})_b^b} &= \frac{1}{2}\Bigl(\hat c^{(1)}(e_a^+)\hat p_b^{-1}(v_a^+) - \hat c^{(1)}(e_a^-)\hat p_b^{-1}(v_a^-)\Bigr) \\
	\widehat{\Delta_{a}^{(v)}(E^{-1})_c^c} &= \frac{1}{2}\Bigl(\hat c^{(1)}(e_a^+)\hat p_c^{-1}(v_a^+) - \hat c^{(1)}(e_a^-)\hat p_c^{-1}(v_a^-)\Bigr) \\
	\widehat{\Delta_{a}^{(v)}(E^{-1})_c^b} &= -\frac{1}{2}\Bigl(\hat s^{(1)}(e_a^+)\hat p_c^{-1}(v_a^+) + \hat s^{(1)}(e_a^-)\hat p_c^{-1}(v_a^-)\Bigr) \\
	\widehat{\Delta_{a}^{(v)}(E^{-1})_b^c} &= \frac{1}{2}\Bigl(\hat s^{(1)}(e_a^+)\hat p_b^{-1}(v_a^+) + \hat s^{(1)}(e_a^-)\hat p_b^{-1}(v_a^-)\Bigr)
	\label{}
\end{align}

\subsubsection*{The operator $\widehat{\Delta_{ab}^{(v)}E_i^c}$}

The components with $a = b$ are given by
\begin{align}
	\widehat{\Delta_{aa}^{(v)}E_a^a} &= \hat p_a(v_a^+) - 2\hat p_a(v) + \hat p_a(v_a^-) \\
	\widehat{\Delta_{aa}^{(v)}E_b^b} &= \hat c^{(1)}(e_a^+)\hat p_b(v_a^+) - 2\hat p_b(v) + \hat c^{(1)}(e_a^-)\hat p_b(v_a^-) \\
	\widehat{\Delta_{aa}^{(v)}E_c^c} &= \hat c^{(1)}(e_a^+)\hat p_c(v_a^+) - 2\hat p_c(v) + \hat c^{(1)}(e_a^-)\hat p_c(v_a^-) \\
	\widehat{\Delta_{aa}^{(v)}E_c^b} &= \hat s^{(1)}(e_a^+)\hat p_b(v_a^+) - \hat s^{(1)}(e_a^-)\hat p_b(v_a^-) \\
	\widehat{\Delta_{aa}^{(v)}E_b^c} &= - \hat s^{(1)}(e_a^+)\hat p_c(v_a^+) + \hat s^{(1)}(e_a^-)\hat p_c(v_a^-)
	\label{}
\end{align}

\noindent The components with $a\neq b$ are given by the expressions below. The operator $\widehat{\Delta_{ab}^{(v)}E_i^c}$ is symmetric in $a$ and $b$, so the components $\widehat{\Delta_{ba}^{(v)}E_i^c}$ are equal to the components given by \Eqs \eqref{eq:Dab-first}--\eqref{eq:Dab-last}.

\begin{align}
	\widehat{\Delta_{ab}^{(v)}E_a^a} &= \frac{1}{8}\biggl[\Bigl(\hat c^{(1)}(e_{ba}^{++}) + \hat c^{(1)}(e_b^+)\Bigr)\hat p_a(v_{ab}^{++}) - \Bigl(\hat c^{(1)}(e_{ba}^{+-}) + \hat c^{(1)}(e_b^-)\Bigr)\hat p_a(v_{ab}^{+-}) \notag \label{eq:Dab-first} \\
	&\qquad\quad - \Bigl(\hat c^{(1)}(e_{ba}^{-+}) + \hat c^{(1)}(e_b^+)\Bigr)\hat p_a(v_{ab}^{-+}) + \Bigl(\hat c^{(1)}(e_{ba}^{--}) + \hat c^{(1)}(e_b^-)\Bigr)\hat p_a(v_{ab}^{--})\biggr] \displaybreak[1] \\[1ex]
	\widehat{\Delta_{ab}^{(v)}E_b^a} &= \frac{1}{8}\biggl[\hat s^{(1)}(e_{ba}^{++})\hat s^{(1)}(e_a^+)\hat p_a(v_{ab}^{++}) + \hat s^{(1)}(e_{ba}^{+-})\hat s^{(1)}(e_a^+)\hat p_a(v_{ab}^{+-}) \notag \\
	&\qquad\quad + \hat s^{(1)}(e_{ba}^{-+})\hat s^{(1)}(e_a^-)\hat p_a(v_{ab}^{-+}) + \hat s^{(1)}(e_{ba}^{--})\hat s^{(1)}(e_a^-)\hat p_a(v_{ab}^{--})\biggr] \displaybreak[1] \\[1ex]
	\widehat{\Delta_{ab}^{(v)}E_c^a} &= \frac{1}{8}\biggl[-\Bigl(\hat s^{(1)}(e_{ba}^{++})\hat c^{(1)}(e_a^+) +\hat s^{(1)}(e_b^+)\Bigr)\hat p_a(v_{ab}^{++}) \notag \\
		&\qquad\quad - \Bigl(\hat s^{(1)}(e_{ba}^{+-})\hat c^{(1)}(e_a^+) + \hat s^{(1)}(e_b^-)\Bigr)\hat p_a(v_{ab}^{+-}) \notag \\
		&\qquad\quad + \Bigl(\hat s^{(1)}(e_{ba}^{-+})\hat c^{(1)}(e_a^-) + \hat s^{(1)}(e_b^+)\Bigr)\hat p_a(v_{ab}^{-+}) \notag \\
	&\qquad\quad + \Bigl(\hat s^{(1)}(e_{ba}^{--})\hat c^{(1)}(e_a^-) + \hat s^{(1)}(e_b^-)\Bigr)\hat p_a(v_{ab}^{--})\biggr] \displaybreak[1] \\[1ex]
	\widehat{\Delta_{ab}^{(v)}E_a^b} &= \frac{1}{8}\biggl[\hat s^{(1)}(e_{ab}^{++})\hat s^{(1)}(e_b^+)\hat p_b(v_{ab}^{++}) + \hat s^{(1)}(e_{ab}^{+-})\hat s^{(1)}(e_b^-)\hat p_b(v_{ab}^{+-}) \notag \\
	&\qquad\quad + \hat s^{(1)}(e_{ab}^{-+})\hat s^{(1)}(e_b^+)\hat p_b(v_{ab}^{-+}) + \hat s^{(1)}(e_{ab}^{--})\hat s^{(1)}(e_b^-)\hat p_b(v_{ab}^{--})\biggr] \displaybreak[1] \\[1ex]
	\widehat{\Delta_{ab}^{(v)}E_b^b} &= \frac{1}{8}\biggl[\Bigl(\hat c^{(1)}(e_{ab}^{++}) + \hat c^{(1)}(e_a^+)\Bigr)\hat p_b(v_{ab}^{++}) - \Bigl(\hat c^{(1)}(e_{ab}^{+-}) + \hat c^{(1)}(e_a^+)\Bigr)\hat p_b(v_{ab}^{+-}) \notag \\
	&\qquad\quad - \Bigl(\hat c^{(1)}(e_{ab}^{-+}) + \hat c^{(1)}(e_a^-)\Bigr)\hat p_b(v_{ab}^{-+}) + \Bigl(\hat c^{(1)}(e_{ab}^{--}) + \hat c^{(1)}(e_a^-)\Bigr)\hat p_b(v_{ab}^{--}) \biggr] \displaybreak[1] \\[1ex]
	\widehat{\Delta_{ab}^{(v)}E_c^b} &= \frac{1}{8}\biggl[\Bigl(\hat s^{(1)}(e_{ab}^{++})\hat c^{(1)}(e_b^+) + \hat s^{(1)}(e_a^+)\Bigr)\hat p_b(v_{ab}^{++}) \notag \\
		&\qquad\quad - \Bigl(\hat s^{(1)}(e_{ab}^{+-})\hat c^{(1)}(e_b^-) + \hat s^{(1)}(e_a^+)\Bigr)\hat p_b(v_{ab}^{+-}) \notag \\
		&\qquad\quad + \Bigl(\hat s^{(1)}(e_{ab}^{-+})\hat c^{(1)}(e_b^+) + \hat s^{(1)}(e_a^-)\Bigr)\hat p_b(v_{ab}^{-+}) \notag \\
	&\qquad\quad - \Bigl(\hat s^{(1)}(e_{ab}^{--})\hat c^{(1)}(e_b^-) + \hat s^{(1)}(e_a^-)\Bigr)\hat p_b(v_{ab}^{--})\biggr] \displaybreak[1] \\[1ex]
	\widehat{\Delta_{ab}^{(v)}E_a^c} &= \frac{1}{8}\biggl[\Bigl(\hat c^{(1)}(e_{ab}^{++})\hat s^{(1)}(e_b^+) + \hat s^{(1)}(e_{ba}^{++})\Bigr)\hat p_c(v_{ab}^{++}) \notag \\
		&\qquad\quad + \Bigl(\hat c^{(1)}(e_{ab}^{+-})\hat s^{(1)}(e_b^-) + \hat s^{(1)}(e_{ba}^{+-})\Bigr)\hat p_c(v_{ab}^{+-}) \notag \\
		&\qquad\quad - \Bigl(\hat c^{(1)}(e_{ab}^{-+})\hat s^{(1)}(e_b^+) + \hat s^{(1)}(e_{ba}^{-+})\Bigr)\hat p_c(v_{ab}^{-+}) \notag \\
	&\qquad\quad - \Bigl(\hat c^{(1)}(e_{ab}^{--})\hat s^{(1)}(e_b^-) + \hat s^{(1)}(e_{ba}^{--})\Bigr)\hat p_c(v_{ab}^{--})\biggr] \displaybreak[1] \\[1ex]
	\widehat{\Delta_{ab}^{(v)}E_b^c} &= \frac{1}{8}\biggl[-\Bigl(\hat c^{(1)}(e_{ba}^{++})\hat s^{(1)}(e_a^+) + \hat s^{(1)}(e_{ab}^{++})\Bigr)\hat p_c(v_{ab}^{++}) \notag \\
		&\qquad\quad + \Bigl(\hat c^{(1)}(e_{ba}^{+-})\hat s^{(1)}(e_a^+) + \hat s^{(1)}(e_{ab}^{+-})\Bigr)\hat p_c(v_{ab}^{+-}) \notag \\
		&\qquad\quad - \Bigl(\hat c^{(1)}(e_{ba}^{-+})\hat s^{(1)}(e_a^-) + \hat s^{(1)}(e_{ab}^{-+})\Bigr)\hat p_c(v_{ab}^{-+}) \notag \\
	&\qquad\quad + \Bigl(\hat c^{(1)}(e_{ba}^{--})\hat s^{(1)}(e_a^-) + \hat s^{(1)}(e_{ab}^{--})\Bigr)\hat p_c(v_{ab}^{--})\biggr] \displaybreak[1] \\[1ex]
	\widehat{\Delta_{ab}^{(v)}E_c^c} &= \frac{1}{8}\biggl[\Bigl(\hat c^{(1)}(e_{ab}^{++})\hat c^{(1)}(e_b^+) + \hat c^{(1)}(e_{ba}^{++})\hat c^{(1)}(e_a^+)\Bigr)\hat p_c(v_{ab}^{++}) \notag \\
		&\qquad\quad - \Bigl(\hat c^{(1)}(e_{ab}^{+-})\hat c^{(1)}(e_b^-) + \hat c^{(1)}(e_{ba}^{+-})\hat c^{(1)}(e_a^+)\Bigr)\hat p_c(v_{ab}^{+-}) \notag \\
		&\qquad\quad - \Bigl(\hat c^{(1)}(e_{ab}^{-+})\hat c^{(1)}(e_b^+) + \hat c^{(1)}(e_{ba}^{-+})\hat c^{(1)}(e_a^-)\Bigr)\hat p_c(v_{ab}^{-+}) \notag \\
	&\qquad\quad + \Bigl(\hat c^{(1)}(e_{ab}^{--})\hat c^{(1)}(e_b^-) + \hat c^{(1)}(e_{ba}^{--})\hat c^{(1)}(e_a^-)\Bigr)\hat p_c(v_{ab}^{--})\biggr]
	\label{eq:Dab-last}
\end{align}

\subsection{Summary}

The expressions given in sections \ref{sec:C_E} and \ref{sec:C_L} provide an explicit definition of a Hamiltonian constraint operator for quantum-reduced loop gravity. The complete constraint operator is given by
\begin{equation}
	{}^R\hat C(N) = \frac{1}{\beta^2}{}^R\hat C_E(N) + \frac{1 + \beta^2}{\beta^2}{}^R\hat C_L(N)
	\label{}
\end{equation}
with the Euclidean and Lorentzian parts of the operator being defined by \Eqs \eqref{eq:CE_R} and \eqref{eq:CE^v_R}, and \Eqs \eqref{eq:CL_R} and \eqref{eq:R_v} respectively. While the operator which has been defined in this section is not symmetric, it is worth keeping in mind that upon symmetrization this operator can be interpreted not only as the Hamiltonian constraint for the vacuum theory, but also as a physical Hamiltonian in the deparametrized context, as indicated by \Eq \eqref{eq:H_phys}.

\section{The one-vertex model}
\label{sec:one-vertex}

\subsection{Hamiltonian operator on single-vertex states}
\label{sec:1v-model}

We will now apply the general expressions developed in the previous section to a simple and concrete example, in which we restrict our attention to reduced spin network states consisting of a single six-valent node. We assume that the spatial manifold has the topology of a three-torus, or is characterized by periodic boundary conditions, such that the spin network graph is formed by three edges, each of which both begins and ends at the single node $v$. Thus, the state space of our one-vertex model is spanned by basis states of the form
\begin{equation}
	\ket{j_xj_yj_z} = \D^{(j_x)}_{j_xj_x}(h_{e_x})_x \D^{(j_y)}_{j_yj_y}(h_{e_y})_y \D^{(j_z)}_{j_zj_z}(h_{e_z})_z.
	\label{eq:1v-basis}
\end{equation}

In order to derive the form of Hamiltonian constraint which governs the dynamics of the states \eqref{eq:1v-basis}, we start with the expressions obtained in the previous section for the action of the Hamiltonian on a general reduced basis state, and identify each node appearing in the expressions with the single node $v$; likewise, each edge aligned with a given coordinate direction $a$ is identified with the edge $e_a$. For the Euclidean part of the Hamiltonian, carrying out these identifications in \Eqs \eqref{eq:CE_R} and \eqref{eq:CE^v_R} yields
\begin{equation}
	{}^R\hat C_E(N) = 4N(v)\biggl[{}^R\hat C_E^{(v, e_x, e_y)} + {}^R\hat C_E^{(v, e_x, e_z)} + {}^R\hat C_E^{(v, e_y, e_z)}\biggr]
	\label{eq:1v-CE}
\end{equation}
with
\begin{equation}
	{}^R\hat C_E^{(v, e_a, e_b)}\ket{j_xj_yj_z} = -\frac{j_aj_b}{\Upsilon_v}\hat c(e_a)\hat s(e_a)\hat c(e_b)\hat s(e_b)\ket{j_xj_yj_z}
	\label{eq:1v-CE^v-unsimplified}
\end{equation}
and the volume eigenvalue is now given by
\begin{equation}
	\Upsilon_v = \sqrt{j_xj_yj_z}.
	\label{}
\end{equation}
The operator \eqref{eq:1v-CE^v-unsimplified} can be written in a slightly more compact form by observing that the operators $\hat c(e)$ and $\hat s(e)$ satisfy the identity
\begin{equation}
	\hat c(e)\hat s(e) = \frac{1}{2}\hat s^{(1)}(e)
	\label{}
\end{equation}
where $\hat s^{(1)}(e)$ is the operator defined by \Eq \eqref{eq:s(1)}, and whose action on the basis states $\ket{j_a} = \D^{(j_a)}_{j_aj_a}(h_{e_a})_a$ reads
\begin{equation}
	\hat s^{(1)}(e_a)\ket{j_a} = \frac{1}{2i}\Bigl(\ket{j_a + 1} - \ket{j_a - 1}\Bigr).
	\label{}
\end{equation}
Hence we have
\begin{equation}
	{}^R\hat C_E^{(v, e_a, e_b)}\ket{j_xj_yj_z} = -\frac{1}{4}\frac{j_aj_b}{\Upsilon_v}\hat s^{(1)}(e_a)\hat s^{(1)}(e_b)\ket{j_xj_yj_z}
	\label{eq:1v-CE^v}
\end{equation}
and using this in \Eq \eqref{eq:1v-CE} yields the expression
\begin{align}
	{}^R\hat C_E(N)\ket{j_xj_yj_z} = -N(v)\Biggl[&\sqrt{\frac{j_xj_y}{j_z}}\hat s^{(1)}(e_x)\hat s^{(1)}(e_y) \notag \\
	&+ \sqrt{\frac{j_xj_z}{j_y}}\hat s^{(1)}(e_x)\hat s^{(1)}(e_z) + \sqrt{\frac{j_yj_z}{j_x}}\hat s^{(1)}(e_y)\hat s^{(1)}(e_z)\Biggr]\ket{j_xj_yj_z}
	\label{eq:1v-CE-explicit}
\end{align}
for the Euclidean Hamiltonian constraint of the one-vertex model.

For the Lorentzian part of the Hamiltonian, we utilize symbolic computer algebra methods (the Python library SymPy \cite{10.7717/peerj-cs.103}) to deal with the lengthy expressions that result when the sums in \Eq \eqref{eq:R_v} are expanded and the expressions given by \Eqs \eqref{eq:Da-first}--\eqref{eq:Dab-last} are substituted for the discretized derivative operators\footnote{
	The code used to perform the calculations is made available at \url{https://github.com/imakinen/QRLG-curvature}.
}. We first generate a fully explicit expression (\ie one written in terms of the spin quantum numbers and reduced holonomy operators, as in \Eq \eqref{eq:CE^v_R} for the Euclidean part) representing the action of the operator ${}^R\hat{\cal R}_v$ on a generic basis state of the reduced Hilbert space. We then specialize this general result to the one-vertex model by performing the identifications specified in the text above \Eq \eqref{eq:1v-CE}. In this way we obtain
\begin{equation}
	{}^R\hat C_L(N)\ket{j_xj_yj_z} = -8N(v)\sum_a \frac{j_a^2}{\Upsilon_v}\biggl(1 - \hat c^{(1)}(e_a) - \frac{1}{2}\bigl[\hat s^{(1)}(e_a)\bigr]^2\biggr)\ket{j_xj_yj_z}
	\label{}
\end{equation}
for the action of the Lorentzian part of the Hamiltonian on the single-vertex state \eqref{eq:1v-basis}. Alternatively, with the help of the identity
\begin{equation}
	1 - \hat c^{(1)}(e) - \frac{1}{2}\bigl[\hat s^{(1)}(e)\bigr]^2 = 2\bigl[\hat s^{(1/2)}(e)\bigr]^4,
	\label{}
\end{equation}
where we now use the notation $\hat s^{(1/2)}(e) \equiv \hat s(e)$ to explicitly indicate the spin of the reduced holonomies entering the operator, the Lorentzian part of the Hamiltonian for the one-vertex model can be expressed in the form
\begin{align}
	&{}^R\hat C_L(N)\ket{j_xj_yj_z} \notag \\
	&= -16N(v)\Biggl[\frac{j_x^{3/2}}{\sqrt{j_yj_z}}\bigl[\hat s^{(1/2)}(e_x)\bigr]^4 + \frac{j_y^{3/2}}{\sqrt{j_xj_z}}\bigl[\hat s^{(1/2)}(e_y)\bigr]^4 + \frac{j_z^{3/2}}{\sqrt{j_xj_y}}\bigl[\hat s^{(1/2)}(e_z)\bigr]^4 \Biggr]\ket{j_xj_yj_z}.
	\label{eq:1v-CL}
\end{align}

\subsection{Analogy with loop quantum cosmology}

It is interesting to compare the structure of the simple model considered above with models describing an anisotropic Bianchi I universe in loop quantum cosmology (see \eg \cite{Ashtekar:2009vc, Chiou:2006qq, Szulc:2008ar, Martin-Benito:2008dfr}), as the Euclidean Hamiltonian constraint given by \Eq \eqref{eq:1v-CE-explicit} bears a certain formal similarity with the Hamiltonian constraint operator considered in these models. The phase space of the classical theory whose quantization leads to the Bianchi I models of loop quantum cosmology is coordinatized by the variables $c_a$ and $p_a$, which correspond essentially to the components of the Ashtekar connection and the densitized triad of a homogeneous spatial geometry, and which are subject to the canonical Poisson brackets
\begin{equation}
	\{c_a, p_b\} \sim \delta_{ab}.
	\label{}
\end{equation}
The Hilbert space of the quantum theory is spanned by the orthonormal basis states
\begin{equation}
	\ket{p_1, p_2, p_3}
	\label{eq:LQC-basis}
\end{equation}
which are eigenstates of the triad operators $\hat p_a$:
\begin{equation}
	\hat p_a\ket{p_1, p_2, p_3} = p_a\ket{p_1, p_2, p_3}.
	\label{eq:LQC-triad}
\end{equation}
As in loop quantum gravity, an operator corresponding to the connection $c_a$ is not available; instead one has the holonomy operators $\widehat{e^{i\mu c_a}}$, whose action on the basis states is defined as
\begin{equation}
	\widehat{e^{i\mu c_1}}\ket{p_1, p_2, p_3} = \ket{p_1 - \mu, p_2, p_3}
	\label{eq:LQC-holonomy}
\end{equation}
and the same for $\widehat{e^{i\mu c_2}}$ and $\widehat{e^{i\mu c_3}}$.

The Hamiltonian constraint of the classical theory is given by
\begin{equation}
	C(N) = -\frac{1}{\beta^2}\frac{N}{\sqrt{p_1p_2p_3}}\bigl(p_1p_2c_1c_2 + p_1p_3c_1c_3 + p_2p_3c_2c_3\bigr).
	\label{eq:C_BI}
\end{equation}
Because of the lack of a well-defined connection operator, the expression \eqref{eq:C_BI} is not directly suitable for quantization. In order to construct a Hamiltonian constraint operator for the quantum theory, one considers a ``polymerized'' version of the classical constraint, which is obtained by making the replacements $c_a \to \sin(\mu_a c_a)/\mu_a$ in \Eq \eqref{eq:C_BI}:
\begin{align}
	C^{(\mu)}(N) = -\frac{1}{\beta^2}N\biggl(&\sqrt{\frac{p_1p_2}{p_3}}\frac{\sin\mu_1c_1}{\mu_1} \frac{\sin\mu_2c_2}{\mu_2} \notag \\
	&+ \sqrt{\frac{p_1p_3}{p_2}}\frac{\sin\mu_1c_1}{\mu_1}\frac{\sin\mu_3c_3}{\mu_3} + \sqrt{\frac{p_2p_3}{p_1}}\frac{\sin\mu_2c_2}{\mu_2}\frac{\sin\mu_3c_3}{\mu_3}\biggr)
	\label{eq:C_BI^mu}
\end{align}
and from which the original constraint \eqref{eq:C_BI} is recovered in the limit $\mu_a \to 0$. The polymerized expression \eqref{eq:C_BI^mu} can be directly quantized by replacing the factors of $\sin\mu_a c_a$ with the operators
\begin{equation}
	\widehat{\sin\mu_a c_a} = \frac{1}{2i}\Bigl(\widehat{e^{i\mu_a c_a}} - \widehat{e^{-i\mu_a c_a}}\Bigr).
	\label{}
\end{equation}
Keeping in mind \Eq \eqref{eq:LQC-holonomy}, it is then immediate to see that the resulting operator is formally identical\footnote{
	However, a notable difference is that in loop quantum cosmology the quantum numbers $p_a$ labeling the basis states \eqref{eq:LQC-basis} are taken to be continuous, unlike the discrete $SU(2)$ quantum numbers $j_a$ in \Eq \eqref{eq:1v-CE-explicit}, although the scalar product between the states \eqref{eq:LQC-basis} is still given by the Kronecker delta instead of the distributional Dirac delta.
} with the operator \eqref{eq:1v-CE-explicit}, provided that the constant value $\mu_a = 1$ is used for the polymerization parameters, and that the factors of $1/\sqrt{p_a}$ in \Eq \eqref{eq:C_BI^mu} are quantized using a Tikhonov-like regularization, as done for example in \cite{Wilson-Ewing:2015xaq}, instead of a regularization where Thiemann identities are used classically to replace the factors of inverse triad with non-singular Poisson brackets.

The formal analogy between the Hamiltonian constraint of Bianchi I loop quantum cosmology and that of the one-vertex model considered in section \ref{sec:1v-model}, if extended to the Lorentzian part of the Hamiltonian, suggests a new, modified approach to the definition of the Hamiltonian constraint operator in loop quantum cosmology. Note that the classical Hamiltonian constraint \eqref{eq:C_BI} of the Bianchi I universe consists only of the Euclidean part, the Lorentzian part being identically zero due to the vanishing Ricci curvature of the homogeneous spatial surfaces. Accordingly, the Hamiltonian constraint operator arising from \eqref{eq:C_BI^mu} also consists of the Euclidean part only. When working with such an operator, one is therefore, in a sense, making the implicit assumption that the Ricci scalar is represented in the quantum theory by an identically vanishing operator.

On the other hand, the form of the Hamiltonian constraint of the one-vertex model, with its Lorentzian part being the non-trivial operator given by \Eq \eqref{eq:1v-CL}, indicates a possible alternative point of view: The Ricci scalar in loop quantum cosmology does not necessarily have to be represented by an operator which is identically zero. Another possibility, which at least on the surface is not inconsistent with the fact that the spatial curvature is classically vanishing, is that the Ricci scalar is represented by a non-trivial operator, which, instead of being identically equal to zero, merely reduces to zero in the limit $\mu\to 0$, and which is formally similar to the operator \eqref{eq:1v-CL} in the same way that the standard Hamiltonian constraint of loop quantum cosmology is similar to the operator \eqref{eq:1v-CE-explicit}.

At the moment we do not have a comprehensive argument which could be regarded as a systematic derivation of the hypothesized operator which is supposed to represent the Ricci scalar in loop quantum cosmology. Nevertheless, we may resort to arguments of a more heuristic nature in order to formulate a preliminary proposal for such an operator. To this end, let us specialize to the case of a homogeneous and isotropic universe, for which the polymerized Euclidean Hamiltonian constraint takes the form
\begin{equation}
	C_E^{(\mu)}(N) = -\frac{3}{\beta^2}N\sqrt p\frac{\sin^2\mu c}{\mu^2}.
	\label{eq:CE-isotropic}
\end{equation}
Heuristically we see that this expression may be obtained from the operator \eqref{eq:1v-CE-explicit} as the result of a process in which the individual holonomy and flux operators become replaced with their classical values\footnote{
	In fact, it has been shown that the expression \eqref{eq:CE-isotropic} can be recovered as the expectation value of the Euclidean constraint operator in a semiclassical state representing a homogeneous and isotropic spatial geometry \cite{Dapor:2017rwv, Assanioussi:2019iye}, although these works used a regularization of the Hamiltonian different from the one considered in the present article.
}; in particular, we suppose that the sine-like operators $\hat s^{(1)}(e_a)$ are replaced with $\sin\mu c$. If we now envision applying an analogous procedure to the operator \eqref{eq:1v-CL}, and assume that the spin-1/2 operators $\hat s^{(1/2)}(e_a)$ get replaced with $\sin(\mu c/2)$, we arrive at
\begin{equation}
	C_L^{(\mu)}(N) = -48\frac{1 + \beta^2}{\beta^2}N\sqrt p\frac{\sin^4(\mu c/2)}{\mu^2}
	\label{eq:CL-isotropic}
\end{equation}
as a potential polymerized expression representing the Lorentzian part of the Hamiltonian. The quantization of this expression according to the one-dimensional version of \Eqs \eqref{eq:LQC-triad} and \eqref{eq:LQC-holonomy} then gives the corresponding quantum operator, in the same way as the quantization of \Eq \eqref{eq:CE-isotropic} gives the standard Hamiltonian constraint operator for homogeneous and isotropic loop quantum cosmology. The complete Hamiltonian constraint of the model would then be given by
\begin{equation}
	\hat C^{(\mu)}(N) = \hat C_E^{(\mu)}(N) + \hat C_L^{(\mu)}(N).
	\label{eq:C^mu-full}
\end{equation}
As the expression \eqref{eq:CL-isotropic} goes to zero in the limit $\mu\to 0$, a tentative physical interpretation of this new term would be that it represents some kind of quantum-gravitational fluctuations of curvature in a universe which is nevertheless characterized by a vanishing spatial curvature at the classical scale.

The modified Hamiltonian constraint for loop quantum cosmology proposed here can be compared with a similar proposal introduced earlier in \cite{Dapor:2017rwv}. In this article the authors worked with Thiemann's well known regularization of the Hamiltonian constraint operator in loop quantum gravity \cite{Thiemann:1996aw}, and obtained a polymerized expression associated with the Lorentzian part of the Hamiltonian by computing the expectation value of the operator with respect to coherent states peaked on homogeneous and isotropic classical data. The resulting polymerized Hamiltonian constraint is
\begin{equation}
	\widetilde C^{(\mu)}(N) = -\frac{3N}{\beta^2\mu^2}\sqrt{p}\Bigl(\sin^2\mu c - (1 + \beta^2)\sin^4\mu c\Bigr).
	\label{}
\end{equation}
In contrast, by applying the identity $\sin^2 x = 4\sin^2(x/2) - 4\sin^4(x/2)$ to \Eq \eqref{eq:CE-isotropic}, the Hamiltonian defined by \Eqs \eqref{eq:CE-isotropic} and \eqref{eq:CL-isotropic} can be written as
\begin{equation}
	C^{(\mu)}(N) = -\frac{12N}{\beta^2\mu^2}\sqrt{p}\Bigl(\sin^2(\mu c/2) + \bigl[4(1 + \beta^2) - 1\bigr]\sin^4(\mu c/2)\Bigr).
	\label{}
\end{equation}
We see that, even though the two expressions are qualitatively similar, they are distinguished from each other by a different relative sign between the quadratic and quartic terms. Therefore our expectation is that the dynamics generated by these two Hamiltonians will be quantitatively different from each other, although an explicit analysis of the dynamics is certainly necessary in order to determine the extent to which the two proposals really differ from each other in terms of their physical content.

To conclude this section, we would like to emphasize that for now the expression \eqref{eq:CL-anisotropic}, as well as its possible generalizations to the anisotropic case, such as
\begin{align}
	C_L^{(\mu)}(N) = -16\frac{1 + \beta^2}{\beta^2}N\Biggl(&\frac{p_1^{3/2}} {\sqrt{p_2p_3}}\frac{\sin^4\bigl(\mu_1c_1/2\bigr)}{\mu_1^2} \notag \\
	&+ \frac{p_2^{3/2}} {\sqrt{p_1p_3}}\frac{\sin^4\bigl(\mu_2c_2/2\bigr)}{\mu_2^2} + \frac{p_3^{3/2}} {\sqrt{p_1p_2}}\frac{\sin^4\bigl(\mu_3c_3/2\bigr)}{\mu_3^2}\Biggr)
	\label{eq:CL-anisotropic}
\end{align}
are best regarded as purely heuristic proposals, motivated by the form of the Lorentzian part of the Hamiltonian of the one-vertex model considered in section \ref{sec:1v-model} together with the formal similarity between the Euclidean Hamiltonian of the one-vertex model and that of anisotropic loop quantum cosmology. These proposals are, for the time being, not supported by detailed calculations which would represent a precise derivation of the proposed expressions. As a further remark, since \Eqs \eqref{eq:CL-isotropic} and \eqref{eq:CL-anisotropic} are not the result of a concrete derivation of a semiclassical effective Hamiltonian, neither from quantum-reduced loop gravity nor from full loop quantum gravity, we are not in a position to make any specific statements on the nature of the polymerization parameters appearing in these expressions. While the structure of the one-vertex model of section \ref{sec:1v-model}, with its states labeled by discrete $SU(2)$ quantum numbers which change in increments of fixed size, is clearly analogous to the $\mu_0$-scheme of loop quantum cosmology, nothing in our current work conclusively rules out the possibility of interpreting \Eq \eqref{eq:CL-isotropic} or \Eq \eqref{eq:CL-anisotropic} in the context of an ``improved dynamics'' scheme \cite{Ashtekar:2006wn}, where the parameters $\mu_a$ are taken to be functions depending on the geometry encoded in the triads $p_a$.

\section{Conclusions}
\label{sec:conclusions}

In this article we considered a concrete implementation of the Hamiltonian constraint operator for quantum-reduced loop gravity. The Euclidean part of our Hamiltonian is similar to operators studied earlier in the literature of loop quantum gravity (\eg in \cite{Alesci:2015wla, Yang:2015zda, Assanioussi:2017tql}) and is distinguished from these operators mainly by the graph-preserving loop assignment adopted in the present work for the holonomy regularizing the Ashtekar curvature. For the Lorentzian part of the Hamiltonian, represented classically by the Ricci scalar of the spatial manifold, we use the curvature operator introduced in \cite{Lewandowski:2021iun} and previously examined in \cite{Lewandowski:2022xox} in the context of quantum-reduced loop gravity. We developed explicit expressions for the action of the Hamiltonian constraint in the reduced spin network basis of quantum-reduced loop gravity.

As a concrete illustration of the expressions giving the action of the Hamiltonian on general reduced basis states, we considered the simple example of reduced spin network states whose graph consists of three orthogonal edges incident on a single six-valent node. (The graph is assumed to be embedded in a spatial manifold having the topology of a three-torus, or being characterized by periodic boundary conditions.) Having established the form of the Hamiltonian constraint which governs the dynamics of such one-vertex states, we observed that there exists a certain formal similarity between the Euclidean part of the Hamiltonian constraint of the one-vertex model, and the Hamiltonian constraint typically used in models of loop quantum cosmology describing an anisotropic Bianchi I universe.

By extending the formal analogy between the Hamiltonian of the one-vertex model and that of Bianchi I loop quantum cosmology to the Lorentzian term, one is led to consider a possible alternative approach to the treatment of the Lorentzian part of the Hamiltonian in loop quantum cosmology. In loop quantum cosmology the Lorentzian part of the Hamiltonian is usually taken to be identically vanishing, on grounds of the classically vanishing curvature of a homogeneous spatial geometry. However, another possibility is that -- just as the quantization of the Euclidean part of the Hamiltonian is based on a ``polymerized'' classical expression, which is not equal to the Hamiltonian constraint of (say) a Bianchi I universe but merely reduces to the latter when the polymerization parameters are taken to zero -- the Lorentzian part may also be represented by a non-trivial polymerized expression, which is not identically vanishing but merely approaches zero in the limit of vanishing polymerization parameters. Physically such a term could be interpreted as describing quantum fluctuations of curvature in a universe whose curvature still remains vanishing at the classical scale.

At the moment we do not have available a concrete calculation or another type of argument which could be seen as a systematic derivation of the operator hypothesized above to represent the Lorentzian part of the Hamiltonian in loop quantum cosmology. On the other hand, it is possible to give a tentative proposal for such an operator by resorting to heuristic arguments based on the form of the Lorentzian part of the Hamiltonian in the one-vertex model of quantum-reduced loop gravity. For the time being, this proposal must be regarded as somewhat speculative, but it could nevertheless be already used as the starting point for constructing modified Hamiltonian operators for models of loop quantum cosmology in order to investigate the possible impact such a modification would have on any physical predictions derived from loop quantum cosmology.

The proposal discussed above could be put on a more solid footing by establishing a methodical derivation of the polymerized expression representing the Lorentzian part of the Hamiltonian from the framework of quantum-reduced loop gravity or full loop quantum gravity. We expect that this could be achieved by evaluating the expectation value of the operator defined by \Eqs \eqref{eq:CL_R} and \eqref{eq:R_v} in coherent states which are peaked on homogeneous/isotropic classical data and interpreting the resulting expression as an effective semiclassical Hamiltonian. (Similar calculations have been performed in \cite{Dapor:2017rwv, Assanioussi:2019iye} to obtain a cosmological effective Hamiltonian corresponding to the Lorentzian part of Thiemann's Hamiltonian in loop quantum gravity.) Even though we consider this to be an important question, we have left it as one to be addressed in future work, mainly due to the potentially high technical complexity of the calculations required to establish a solid result on the semiclassical expectation values of the Hamiltonian. Having a derivation of this kind would be particularly important in order to confirm whether the dependence on the polymerization parameters in expressions like \eqref{eq:CL-isotropic} and \eqref{eq:CL-anisotropic} is indeed such that these expressions reduce to zero in the limit of vanishing polymerization parameters, which is necessary for a consistent interpretation of these expressions as representing the curvature of a classically homogeneous spatial geometry.

Another essential topic for future work is to study explicit examples of the dynamics generated by the Hamiltonian operator considered in this article in order to extract concrete information about the physics contained in the new Hamiltonian, and to determine whether this new proposal carries any genuinely new physical content. This question can be examined on several different levels: semiclassical effective dynamics, the loop quantum cosmology dynamics arising from the modified Hamiltonian \eqref{eq:C^mu-full} or its anisotropic generalizations, and the quantum dynamics of states in quantum-reduced loop gravity defined on the one-vertex graph. We believe all of these approaches to be practically accessible at least in terms of numerical simulations.

\subsection*{Acknowledgments}

The author thanks Mehdi Assanioussi for helpful comments on the manuscript. This work was funded by National Science Centre, Poland through grant no.~2022\slash 44\slash C\slash ST2\slash 00023. For the purpose of open access, the author has applied a CC BY 4.0 public copyright license to any author accepted manuscript (AAM) version arising from this submission.

\appendix

\section{Eigenstates of the angular momentum operator}
\label{sec:J}

The states $\ket{jm}_i$ $(i = x, y, z)$, which enter the definition of the basis states \eqref{eq:basis}, are defined by the eigenvalue equations
\begin{align}
	J^2\ket{jm}_i &= j(j+1)\ket{jm}_i \\
	J_i\ket{jm}_i &= m\ket{jm}_i
	\label{}
\end{align}
Given the standard basis $\ket{jm} \equiv \ket{jm}_z$, in which $J_z$ is diagonal, the states $\ket{jm}_i$ for $i = x$ or $y$ can be constructed as
\begin{equation}
	\ket{jm}_i = D^{(j)}(g_i)\ket{jm}
	\label{eq:jm_i}
\end{equation}
where $g_i$ is an $SU(2)$ group element corresponding to a rotation which rotates the $z$-axis into the $i$-axis. This requirement is of course not sufficient to uniquely determine the group element $g_i$. In order to remove the ambiguity, we require that the rotation encoded in $g_i$ corresponds to a cyclic permutation of the coordinate axis, \ie the axes $(x, y, z)$ are mapped into $(z, x, y)$ and $(y, z, x)$ under the rotations corresponding to $g_x$ and $g_y$ respectively. Under this choice, the group elements $g_x$ and $g_y$ are explicitly given by
\begin{align}
	g_x &= g_{\hat e_y}(\pi/2) g_{\hat e_z}(\pi/2)
	\label{eq:g_x} \\
	g_y &= g_{\hat e_x}(-\pi/2) g_{\hat e_z}(-\pi/2)
	\label{eq:g_y}
\end{align}
where
\begin{equation}
	g_{\vec n}(\alpha) = e^{-i\alpha\vec n\cdot\vec\sigma/2}
	\label{}
\end{equation}
denotes the group element corresponding to a rotation by the angle $\alpha$ around the axis defined by the unit vector $\vec n$.

In calculations involving the reduced holonomy operator, it is often necessary to perform a change of basis between the standard basis $\ket{jm}$ and the bases $\ket{jm}_i$ for $i=x$ and $y$. For the work presented in this article, where the holonomy operators entering the Euclidean part of the Hamiltonian are taken in the fundamental representation, it is sufficient to know how these bases are related to each other for $j = 1/2$. By constructing the matrices \eqref{eq:g_x} and \eqref{eq:g_y} and applying \Eq \eqref{eq:jm_i} for the states $\ket{\pm} \equiv \ket{\tfrac{1}{2}, \pm\tfrac{1}{2}}$ in the $j = 1/2$ representation, we find
\begin{align}
	\ket{+}_x &= \frac{1}{\sqrt 2}e^{-i\pi/4}\ket{+} + \frac{1}{\sqrt 2}e^{-i\pi/4}\ket{-} \\
	\ket{-}_x &= -\frac{1}{\sqrt 2}e^{i\pi/4}\ket{+} + \frac{1}{\sqrt 2}e^{i\pi/4}\ket{-}
	\label{}
\end{align}
and
\begin{align}
	\ket{+}_y &= \frac{1}{\sqrt 2}e^{i\pi/4}\ket{+} - \frac{1}{\sqrt 2}e^{-i\pi/4}\ket{-} \\
	\ket{-}_y &= \frac{1}{\sqrt 2}e^{i\pi/4}\ket{+} + \frac{1}{\sqrt 2}e^{-i\pi/4}\ket{-}
	\label{}
\end{align}
The inverse change of basis is given by
\begin{align}
	\ket{+} &= \frac{1}{\sqrt 2}e^{i\pi/4}\ket{+}_x - \frac{1}{\sqrt 2}e^{-i\pi/4}\ket{-}_x
	\label{eq:+(x)} \\
	\ket{-} &= \frac{1}{\sqrt 2}e^{i\pi/4}\ket{+}_x + \frac{1}{\sqrt 2}e^{-i\pi/4}\ket{-}_x
	\label{eq:-(x)}
\end{align}
and
\begin{align}
	\ket{+} &= \frac{1}{\sqrt 2}e^{-i\pi/4}\ket{+}_y + \frac{1}{\sqrt 2}e^{-i\pi/4}\ket{-}_y
	\label{eq:+(y)} \\
	\ket{-} &= -\frac{1}{\sqrt 2}e^{i\pi/4}\ket{+}_y + \frac{1}{\sqrt 2}e^{i\pi/4}\ket{-}_y
	\label{eq:-(y)}
\end{align}

\printbibliography

\end{document}